\documentclass[11pt,3p,review,authoryear]{elsarticle}

\usepackage{amsmath}
\usepackage[singlelinecheck=false]{caption}
\usepackage{multirow}
\usepackage{adjustbox}
\usepackage{pdflscape}
\usepackage[colorlinks,allcolors=red]{hyperref}
\usepackage{hhline}
\usepackage{mathtools}
\usepackage{booktabs}

\newcommand{\thicker}[1]{%
  \textbf{#1}\llap{\textbf{\kern.1em #1}}%
}

\begin{document}

\begin{frontmatter}

\title{Modelling and forecasting energy market volatility using GARCH and machine learning approach}

\author[inst1]{Seulki Chung}

\affiliation[inst1]{organization={GSEFM, Department of Empirical Economics, Technische Universität Darmstadt},
            addressline={Karolinenpl.5}, 
            city={Darmstadt},
            postcode={64289}, 
            country={Germany}}



\begin{abstract}
This paper presents a comparative analysis of univariate and multivariate GARCH-family models and machine learning algorithms in modeling and forecasting the volatility of major energy commodities: crude oil, gasoline, heating oil, and natural gas. It uses a comprehensive dataset incorporating financial, macroeconomic, and environmental variables to assess predictive performance and discusses volatility persistence and transmission across these commodities. Aspects of volatility persistence and transmission, traditionally examined by GARCH-class models, are jointly explored using the SHAP (Shapley Additive exPlanations) method. The findings reveal that machine learning models demonstrate superior out-of-sample forecasting performance compared to traditional GARCH models. Machine learning models tend to underpredict, while GARCH models tend to overpredict energy market volatility, suggesting a hybrid use of both types of models. There is volatility transmission from crude oil to the gasoline and heating oil markets. The volatility transmission in the natural gas market is less prevalent.
\linebreak
\end{abstract}


\end{frontmatter}
\newpage
\section{Introduction}

Energy serves as a fundamental building block for driving economic development. Fluctuations in its pricing significantly influence the flow and allocation of resources within the energy market, yielding substantial economic impact. (\citet{kaufmann2020, hamilton1983}). This inherent volatility in energy markets, depicted by fluctuations in the prices of energy commodities such as crude oil, has implications for economic stability, investment strategies, and policymaking decisions (\citet{pindyck1999}). High volatility can discourage investment in fixed capital due to price uncertainty and encourage firms to protect their assets against price risk. Additionally, high volatility can lead to greater demand for storage, resulting in higher spot prices and convenience yield. Understanding changes in volatility can help explain shifts in other economic variables (\citet{henriques2011,karali2014,pindyck2004b}). Therefore, accurately forecasting this volatility is important as it carries direct implications for hedging and derivatives trading. Furthermore, time-varying volatility is subject to heteroskedasticity in the data, resulting in biased standard error estimates and invalid statistical inference (\citet{efimova2014}). However, energy market volatility is determined by various factors, including geopolitical events, supply-demand imbalances, regulatory changes, and macroeconomic factors, and is thus difficult to model and predict. 

Historically, the volatility of energy prices, particularly crude oil, has been an important point of academic interest due to its wide-ranging implications on economic activity, and the literature on energy market volatility has evolved substantially over the past few decades. Traditional econometric models, especially the Generalized Autoregressive Conditional Heteroskedasticity (GARCH) model introduced by \citet{engle1982} and further developed by \citet{bollerslev1986}, have been extensively employed to model the conditional variance characteristic of time series data. Their variants such as the Exponential GARCH (EGARCH) introduced by \citet{nelson1991} and the Glosten-Jagannathan-Runkle GARCH (GJR-GARCH) proposed by \citet{glosten1993}, have also been used to capture the asymmetric effects of shocks on market volatility, a feature particularly pronounced in energy markets (\citet{sadorsky1999,reboredo2011}). However, the focus was on univariate GARCH models, and \citet{lin2001,morana2001} are empirical examples that apply univariate GARCH models to energy data. A drawback of univariate models is that they cannot reveal the relationships between the energy markets. On the other hand, multivariate GARCH models help investigate correlations and volatility spillover among energy markets and hedging strategies (\citet{haigh2002,sadorsky2006,sadorsky2012}). Thus, more recently, the studies by \citet{karali2014,efimova2014,wang2012} demonstrate the applicability of univariate and multivariate GARCH models in forecasting the volatility of energy commodities and assessing market risk.

Despite their widespread adoption, GARCH-family models encounter some limitations. \citet{dritsaki2017} points out that GARCH models assume only the size of the return of the conditional variance is defined, without considering the positivity or negativity of volatility's return, which is unpredictable. Another crucial limitation is the non-negativity of parameters to ensure the positivity of the conditional variance. These limitations make the estimation of GARCH models difficult. Therefore they have driven the exploration of alternative methodologies, with recent scholarly interests in machine learning techniques, noted for their capability to model nonlinear relationships without pre-specified assumptions about the functional form and the statistical distribution of parameters (\citet{ghoddusi2019}). Furthermore, machine learning approaches have gained traction for their ability to extract and interpret complex patterns from vast datasets (\citet{huang2017}). In energy economics, the adoption of machine learning has been slower, yet an emerging body of literature suggests significant potential (\citet{ghoddusi2019,lu2021}). While GARCH models have long been the standard for measuring time-varying volatility and assessing risk, they haven't been thoroughly compared with machine learning models, which bring new perspectives on energy market dynamics based on advancements in computing and data availability.

This paper's contribution to the extant literature is fourfold. Firstly, it extends previous analyses by conducting a comparative study of univariate and multivariate GARCH models applied to four key energy commodities, addressing the gap left by prior research mainly focusing on bivariate or trivariate GARCH models. Secondly, it undertakes a detailed comparative analysis of GARCH-family models and selected machine learning algorithms, evaluating their efficacy in modeling and forecasting energy market volatility. This comparison is important, given each approach's distinct methodological underpinnings and assumptions. Thirdly, the study explores volatility transmission and spillover effects, traditionally examined by multivariate GARCH models. Extending this analysis to machine learning frameworks, the research provides new insights into the interconnectedness and systemic risk pervading global energy markets. Finally, this investigation is distinguished by its comprehensive incorporation of a wide array of explanatory variables from various market-specific domains such as finance, macroeconomics, economic policy uncertainty, and temperature conditions, thus significantly increasing the forecasting accuracy and robustness of its predictive models. This approach addresses the academic recommendations for models that better reflect the complex factors affecting energy prices, as noted in established studies by \citet{kilian2009} and \citet{hamilton2009}, and further supported by recent work from \citet{karali2014} and \citet{reboredo2016}.

The remainder of the paper is as follows: Section 2 explains the data used. Section 3 describes the models and performance evaluation metrics and outlines the research methodology. Section 4 presents the in-sample estimation, the out-of-sample prediction results, and the results of the SHAP method. Section 5 concludes.

\section{Data}

This study employs a comprehensive dataset to model and forecast daily volatility in energy markets, focusing on four energy commodities from 2006 to 2023: Crude Oil (CO), Gasoline (G), Heating Oil (HO), and Natural Gas (NG). Specifically, I use the West Texas Intermediate crude oil price at Cushing, Oklahoma, the conventional gasoline price at New York Harbor, the heating oil price at New York Harbor, and the natural gas price at Henry Hub. These commodities' spot prices and inventory levels are obtained from the U.S. Energy Information Administration (EIA), which provides weekly updates and historical data. The selection of these commodities is motivated by their representative nature of the broader energy sector and data availability. The theory of storage implies increasing price volatility when inventory levels decline, and \citet{suenaga2011} argue that declining inventories with inelastic demand let even small shocks to demand or supply bring about large price fluctuations. \citet{ng1994} and \citet{pindyck2004a} find empirical evidence that inventory levels affect the price volatility of metal and heating oil, respectively. 

Complementing the commodity-specific data, a set of financial and macroeconomic variables obtained from the Federal Reserve Economic Data (FRED) database is incorporated. These variables include the federal funds rate (FFR), the S\&P 500 index, the U.S. dollar index, the consumer price index, the producer price index, the industrial production index, the unemployment rate, and the term spread between 10-year and 3-month Treasury securities. The selection of these variables is grounded in the literature's recognition of their influence on energy markets. For instance, \citet{slade2006} find that the industrial production index as a measure of business cycles influences the price fluctuations in the metal markets. The federal funds rate impacts the carrying cost of inventories, thereby storage decisions and prices, while the S\&P 500 index reflects broader market sentiment and economic health (\citet{sadorsky1999}). The U.S. Dollar Index is relevant for commodities priced in dollars, influencing international demand (\citet{karali2014}). The consumer and producer price indexes offer insights into inflationary pressures. Including the unemployment rate and term spreads provides additional economic context, reflecting labor market conditions and interest rate expectations, respectively.

This study also incorporates the Global Economic Policy Uncertainty (GEPU) Index. This index can be accessed through its dedicated public website, offering data reflecting the uncertainty in global economic policy, compiled from news coverage frequency, tax code provisions, and disagreement among economic forecasters. Economic policy uncertainty can lead to fluctuations in energy prices by affecting investor sentiment, market stability, and future energy demand projections. Specifically, higher levels of uncertainty can deter investment in energy infrastructure and projects, leading to supply constraints and price volatility. Furthermore, uncertainty in economic policies can impact energy consumption patterns and prices, as businesses and consumers alter their spending and energy use in response to changing economic outlooks.

Moreover, this study integrates temperature data from the NASA Goddard Institute for Space Studies (GISS), focusing on the Combined Land-Surface Air and Sea-Surface Water Temperature Anomalies (Land-Ocean Temperature Index, L-OTI). This dataset represents deviations from the mean temperatures of the 1951-1980 period, encapsulating global and regional climatic trends. The rationale for including temperature anomalies stems from the growing recognition of climate factors' impact on energy consumption patterns, especially in heating and cooling demand (\citet{decian2019,li2021}). Additionally, temperature variations are linked to renewable energy outputs, such as solar and wind power, thus affecting overall energy market dynamics.

\begin{table}[!ht]
    \caption{Overview of predictors}
    \resizebox{\textwidth}{!}{\begin{tabular}{lrrrrrrrr}
\toprule
           &  Mean (\%) &  Max Value &  Min Value &  Std Dev (\%) & Jarque-Bera Stat &              ADF Stat &               PP Stat &               KPSS Stat \\
\midrule
       Crude Oil Spot &    -0.035 &      0.531 &     -3.020 &        5.653 &            $<$0.01*** & -12.552*** & -52.046*** &  0.097 \\
        Gasoline Spot &     0.083 &      0.622 &     -0.362 &        3.938 &            $<$0.01*** & -13.742*** &  -65.682*** &  0.030\\
     Heating Oil Spot &     0.037 &      0.280 &     -0.200 &        2.438 &            $<$0.01*** & -67.770*** & -67.770*** &  0.067 \\
     Natural Gas Spot &     0.119 &      1.108 &     -0.641 &        5.629 &            $<$0.01*** & -16.178*** & -64.618*** &    0.320\\
  Crude Oil Inventory &    -0.005 &      0.012 &     -0.010 &        0.121 &            $<$0.01*** & -9.351*** & -42.245*** &   0.338\\
   Gasoline Inventory &     0.020 &      0.062 &     -0.043 &        0.712 &            $<$0.01*** & -15.347*** & -34.529*** &   0.162\\
Heating Oil Inventory &     0.000 &      0.063 &     -0.026 &        0.498 &            $<$0.01*** & -11.584*** & -41.857*** &  0.022 \\
Natural Gas Inventory &     0.011 &      0.051 &     -0.091 &        1.015 &           $<$0.01*** & -9.164*** & -55.067*** &  0.012\\
         Dollar Index &     0.004 &      0.019 &     -0.025 &        0.347 &            $<$0.01*** & -16.874*** & -64.911*** &   0.142\\
                  FFR &     0.493 &      3.125 &     -0.773 &       11.196 &            $<$0.01*** & -13.881*** & -82.913*** &   0.247\\
                SP500 &     0.037 &      0.116 &     -0.120 &        1.248 &           $<$0.01*** & -15.310*** & -77.499*** &   0.132\\
          Term Spread &    -0.475 &      5.000 &    -13.000 &       43.439 &           $<$0.01*** & -14.500*** & -64.530*** &  0.044\\
 Consumer Price Index &     0.010 &      0.002 &     -0.002 &        0.019 &            $<$0.01*** & -6.593*** & -62.501*** &   0.332\\
Industrial Production &     0.001 &      0.008 &     -0.016 &        0.079 &           $<$0.01*** & -10.571*** & -46.803*** &  0.048 \\
 Producer Price Index &     0.009 &      0.004 &     -0.006 &        0.075 &            $<$0.01*** & -5.927*** & -51.845*** &   0.130 \\
    Unemployment Rate &    -0.004 &      0.180 &     -0.042 &        0.617 &            $<$0.01*** & -9.209*** & -44.283*** &  0.080\\
                 GEPU &     0.036 &      0.076 &     -0.056 &        1.089 &            $<$0.01*** & -12.540*** & -42.259*** &  0.013 \\
Temperature & 0.022 & 0.100 & -0.072 & 1.013 & $<$0.01*** & -13.017*** & -40.565*** & 0.009 \\
\hline
\vspace{-1cm}
\end{tabular}}
    \caption*{The table presents a list of energy commodities and predictive variables. It includes descriptive statistics and test statistics to check normality and stationarity. \\
    *** denotes rejection of the null hypothesis at 1\% significance level.}
\end{table}

Table 1 presents the descriptive statistics of energy returns, obtained by calculating the first-order difference of logarithmic price, and logarithmic or simple first-order differenced series of other predictors.

The different return series exhibit a relatively broad range of values, reflecting the diverse nature and volatility inherent in these markets. Notably, natural gas spot prices show a higher mean return of 0.119\%, compared to a slight negative mean return for Crude Oil Spot at -0.035\%. The mean return of natural gas prices has been higher than that of crude oil prices from 2006 to 2023, primarily due to the distinct supply and demand dynamics affecting each market, including technological advancements in natural gas extraction and shifts towards cleaner energy sources.



The Jarque-Bera statistics, indicating non-normality at the null hypothesis, suggest that the return distributions deviate significantly from the normal distribution. The stationarity of the series is confirmed by the ADF and PP tests, where the ADF and PP statistics reject the null hypothesis of a unit root. The KPSS test shows that the null hypothesis of stationarity cannot be rejected for all the variables listed here, indicating that the energy return process is stationary.

\begin{figure}[!ht]
\caption{Energy commodities: Returns}
\includegraphics[width=\textwidth]{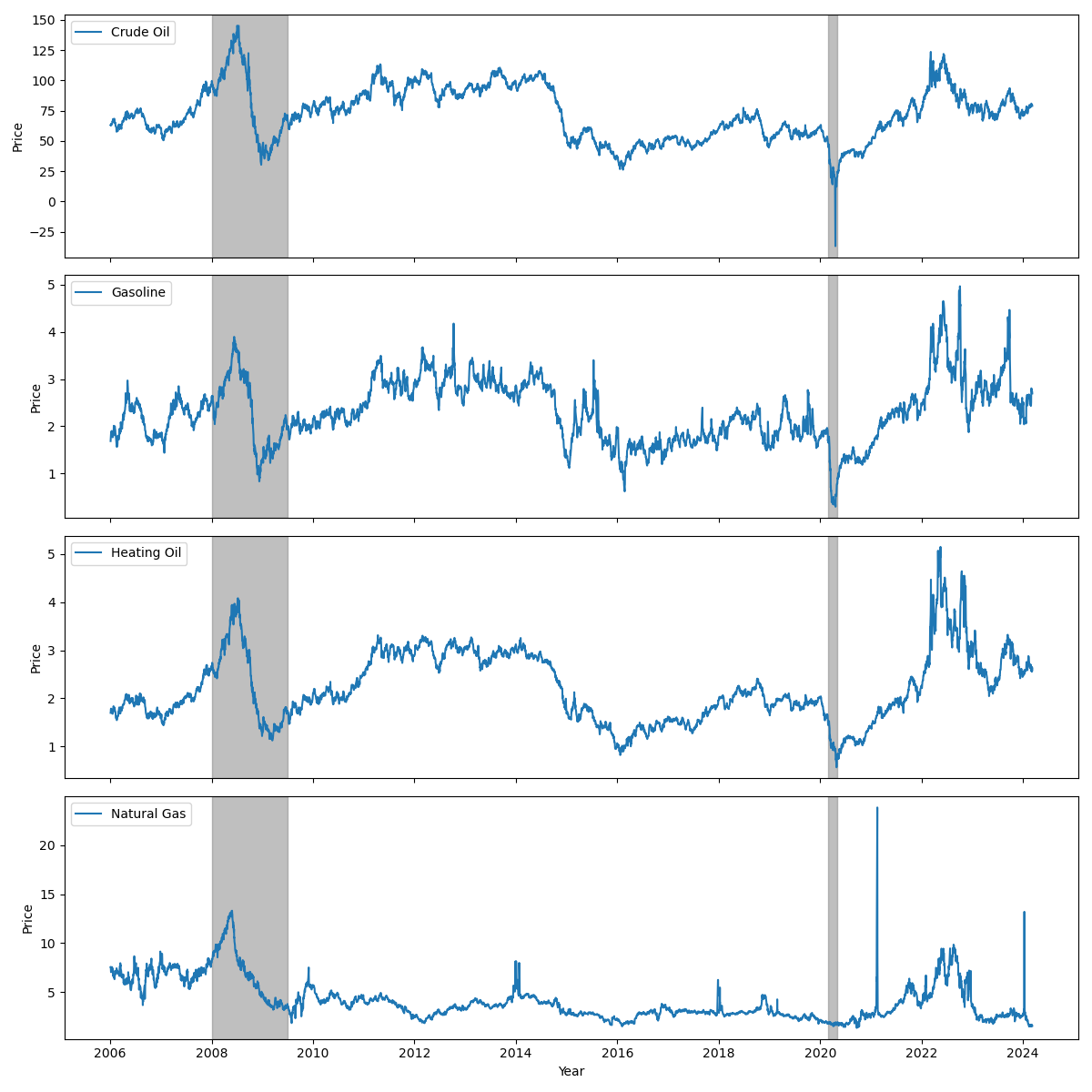}\\
\caption*{The figure illustrates the development of the spot prices of crude oil, gasoline, heating oil, and natural gas from January 2006 to December 2023. The grey-shaded areas depict NBER recession months.}
\end{figure}

Figure 1 provides an overview of the development of the spot prices of four energy commodities: crude oil, gasoline, heating oil, and natural gas, spanning from January 2006 through December 2023. The graph displays notable volatility in crude oil returns, with a significant downturn observed around 2008, corresponding with the global financial crisis, as documented by \citet{hamilton2009}. This period, highlighted by a grey shaded area, signifies a pronounced market reaction to economic contraction, emphasizing crude oil's sensitivity to global economic health. Post-2008, the market exhibits recovery, albeit with persistent fluctuations indicative of geopolitical tensions and supply-demand dynamics. Another sharp decline in 2020, tied to the onset of the COVID-19 pandemic, shows the vulnerability of crude oil markets to sudden demand shocks (\citet{kilianpark2009}). Given their direct supply chain linkage, the gasoline market trends are similar to those of crude oil, with volatility particularly pronounced during the 2008 financial crisis and the 2020 pandemic. The subsequent price spikes in 2020 and beyond hint at recovery phases, yet indicate the market's susceptibility to refinery outputs, seasonal consumption patterns, and crude oil price changes. The correlation between gasoline and crude oil prices, examined in studies such as \citet{brown2002}, is visually corroborated, reflecting the integrated nature of energy commodities. Heating oil returns display relative stability outside the recessionary periods, with less pronounced volatility than crude oil and gasoline. This stability may be attributed to the consistent demand for heating, especially in colder regions, albeit with notable dips during economic downturns. The graph aligns with the work of EIA reports, suggesting that heating oil's price dynamics are influenced by both crude oil prices and seasonal demand fluctuations. The natural gas segment presents a distinct pattern, with a massive spike in volatility around 2008, followed by significant fluctuations thereafter. The volatility peaks, especially in the early 2020s, may reflect the start of the Russia-Ukraine War.

\begin{figure}[!ht]
\caption{Energy commodities: Volatilities}
\includegraphics[width=\textwidth]{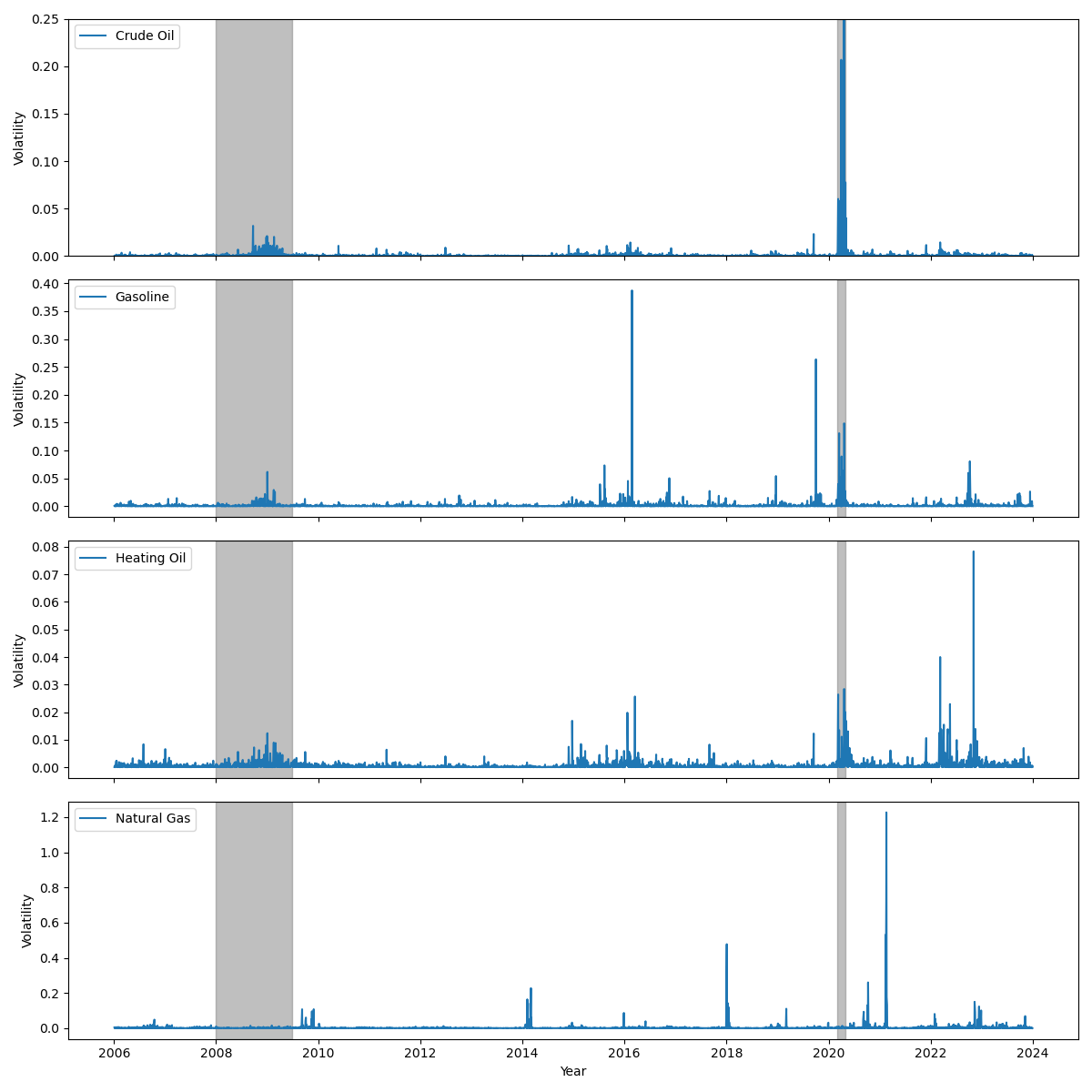}\\
\caption*{The figure illustrates the volatilities of crude oil, gasoline, heating oil, and natural gas from January 2006 to December 2023. The grey-shaded areas depict NBER recession months.}
\end{figure}

The graphical representation of the volatility of energy commodities is shown in Figure 2, where the squared return is taken as the proxy for actual volatility. Crude oil displays periods of heightened volatility, particularly pronounced in 2020. The spike in volatility during these periods can be attributed to sudden drops in demand and supply-side shocks due to the pandemic. This volatility pattern aligns with the findings from \citet{kilian2009}, which emphasizes the impact of external economic factors on oil prices. The subsequent stabilization and sporadic peaks post-2020 suggest a market adapting to new supply-demand dynamics and geopolitical influences. Gasoline volatility, except for some spikes in and before 2020, generally follows the trends observed in crude oil. Heating Oil displays a relatively steady volatility pattern, largely attributed to seasonal demand variations, especially in regions dependent on heating oil for winter heating. Natural Gas volatility is characterized by distinct peaks, diverging from the patterns observed in liquid fuels. The volatility spikes, especially in the late 2010s and early 2020s, may indicate the evolving natural gas market, influenced by technological advancements in extraction, changing consumption patterns, and the increasing global trade in liquefied natural gas (LNG). These trends highlight the unique market dynamics of natural gas, which is less directly tied to oil prices but increasingly subject to global market forces and environmental policies.

\section{Model}

\subsection{GARCH}

\subsubsection{Univariate GARCH}

The baseline model for the analysis is the Generalized Autoregressive Conditional Heteroskedasticity (GARCH) model, as proposed by \citet{bollerslev1986}. The GARCH(1,1) model, an extension of Engle’s ARCH model (\citet{engle1982}), is formulated as follows:

\[
r_t = \mu + \epsilon_t,
\]
\[
\epsilon_t = \sigma_t z_t,
\]
\[
\sigma^2_t = \omega + \alpha \epsilon^2_{t-1} + \beta \sigma^2_{t-1},
\]

where \(r_t\) represents the logarithmic returns of the energy commodities at time \(t\), \(\mu\) is the mean return, \(\epsilon_t\) is the residual at time \(t\), \(\sigma_t^2\) is the conditional variance, \(z_t\) is an i.i.d. shock term with mean 0 and standard deviation 1, \(\omega\) is a constant term, and \(\alpha\) and \(\beta\) are parameters representing the short-term persistence of shocks and the long-term variance persistence, respectively.

To extend the standard GARCH(1,1) model for incorporating external regressors, the conditional variance equation is augmented as follows:

\[
\sigma^2_t = \omega + \alpha \epsilon^2_{t-1} + \beta \sigma^2_{t-1} + \sum_{i=1}^{m} \gamma_i X_{i,t},
\]

where \(X_{i,t}\) represents the \(i\)-th external regressor at time \(t\), and \(\gamma_i\) are the coefficients quantifying the impact of these external factors on the conditional variance.

In addition to the standard GARCH(1,1) framework, the Exponential GARCH (EGARCH) model and the Glosten-Jagannathan-Runkle GARCH (GJR-GARCH) model are employed to capture the asymmetries and leverage effects prevalent in energy markets. The EGARCH model, proposed by \citet{nelson1991}, is specified as:

\[
\log(\sigma^2_t) = \omega + \alpha (|z_{t-1}| - E|z_{t-1}|) + \gamma z_{t-1} + \beta \log(\sigma^2_{t-1}),
\]

\noindent and the GJR-GARCH model, following \citet{glosten1993}, is represented by:

\[
\sigma^2_t = \omega + (\alpha + \gamma I_{t-1}) \epsilon^2_{t-1} + \beta \sigma^2_{t-1},
\]

\noindent where \(I_{t-1}\) is an indicator function that takes the value of 1 if \(\epsilon_{t-1} < 0\) and 0 otherwise, allowing the model to differentiate between positive and negative shocks to the system.

\subsubsection{Multivariate GARCH}
For the multivariate component, the study incorporates the GARCH-BEKK model from \citet{engle1995}, a prevalent form of the multivariate GARCH model that provides a systematic approach to modeling volatility spillovers across multiple commodities. The BEKK model is specified as follows:

\[
H_t = C'C + A'\epsilon_{t-1}\epsilon_{t-1}'A + B'H_{t-1}B,
\]

\noindent where \(H_t\) is the conditional covariance matrix of the returns, \(C\) is a lower triangular matrix, and \(A\) and \(B\) are matrices that capture the responses of the current conditional variances to past squared innovations and past conditional variances, respectively. This formulation ensures the positive definiteness of the conditional covariance matrix, which is necessary for volatility modeling.

The stationarity condition for the BEKK process involves the spectral radius of the matrices \(A\) and \(B\). Specifically, the process is considered stationary if the eigenvalues of the matrix \(M = A \otimes A + B \otimes B\) (where \(\otimes\) denotes the Kronecker product) satisfy the condition:

\[
\rho(M) < 1,
\]

\noindent where \(\rho(M)\) denotes the spectral radius of \(M\), which is the largest absolute value among the eigenvalues of \(M\). If this condition is met, the effects of shocks to the system diminish over time, ensuring stable long-term behavior of the conditional covariance matrix \(H_t\). This stationarity condition must be met for the theoretical consistency and empirical applicability of the model in volatility analysis.

Each of these models rests on distinct assumptions about market behavior and statistical properties. The GARCH(1,1) model assumes symmetric effects of shocks on volatility, while the EGARCH and GJR-GARCH models allow for asymmetry, capturing the leverage effect observed in many financial series. The multivariate BEKK model extends this framework to a system of interrelated time series, facilitating the examination of volatility spillover effects among different energy commodities.

\subsection{Machine learning models}

This section outlines the machine learning models used to forecast energy market volatility. Specifically, standard and penalized linear regression models, Bayesian ridge, decision trees, random forests, XGBoost, support vector machines (SVM), K-nearest neighbors (KNN), and multi-layer perceptron (MLP) are briefly discussed with the primary goal of building a comprehensive framework for modeling and predicting energy market volatility.

Linear regression serves as a benchmark in our analysis. The model is defined as:

\[
\sigma = X\beta + \epsilon,
\]

\noindent where \(\sigma\) represents the volatility measures of four energy commodities, \(X\) denotes the matrix of explanatory variables, \(\beta\) symbolizes the coefficients, and \(\epsilon\) is the error term. Despite its simplicity, linear regression provides a baseline for performance comparison.

Regularization techniques such as ridge and lasso regression modify the linear regression framework to address multicollinearity and overfitting, especially pertinent in high-dimensional datasets. The ridge regression introduces an \(L_2\) penalty term:

\[
\min_{\beta} \left\{ ||\Sigma - X\beta||_2^2 + \lambda ||\beta||_2^2 \right\},
\]

while lasso regression employs an $L_1$ penalty:

\[
\min_{\beta} \left\{ ||\Sigma - X\beta||_2^2 + \lambda ||\beta||_1 \right\},
\]

\noindent where \(\lambda\) is the regularization parameter. Elastic net, a hybrid approach, combines both \(L_1\) and \(L_2\) penalties, optimizing the balance between ridge and lasso characteristics. Bayesian ridge extends these concepts within a Bayesian framework, offering probabilistic interpretations of regression coefficients.

Decision trees decompose the data space into a series of decisions, based on which the response variable is predicted. While individual trees are prone to overfitting, ensemble methods like random forests and XGBoost mitigate this risk. Random forests construct multiple decision trees during training and output the average prediction, enhancing model robustness. XGBoost, an implementation of gradient boosted trees, optimizes the model using gradient descent, addressing shortcomings of traditional boosting techniques. XGBoost functions by sequentially adding decision trees, where each new tree corrects errors made by the previously added trees. It uses a gradient boosting framework, meaning it applies gradient descent to minimize the loss when adding new models. Unlike simpler models, XGBoost incorporates regularization (L1 and L2), which helps reduce overfitting by penalizing complex models. It efficiently handles sparse data and missing values by assessing whether a split on missing values will improve the model's performance.

The main differences between XGBoost and Random Forest lie in their approach and capabilities. Random Forest builds many decision trees in parallel and averages their predictions. Each tree in a Random Forest is built from a random sample of the data and uses a random subset of features for splitting nodes. This randomness helps in reducing overfitting, making the model more robust. In contrast, XGBoost builds one tree at a time, where each new tree corrects errors made by previously built trees, and it explicitly includes regularization terms in its optimization problem, which can lead to better generalization on unseen data. Additionally, while Random Forest can be parallelized across trees, XGBoost also parallelizes the construction of each tree, making it often faster and more efficient on large datasets.

Support vector machines or support vector regression constructs a hyperplane in a high-dimensional space to predict continuous values, optimizing the margin while penalizing points that fall outside the error margin. K-nearest neighbors, in contrast, predict the response based on the average of its \(k\) nearest neighbors in the variable space, offering a non-parametric approach to forecasting.

The multi-layer perceptron, a class of feedforward artificial neural networks, consists of multiple layers of nodes in a directed graph, with each layer fully connected to the subsequent one. The MLP model captures complex nonlinear relationships through its network structure and activation functions:

\[
y = f(W_n \cdot f(...f(W_2 \cdot f(W_1 \cdot X + b_1) + b_2)... + b_n),
\]

\noindent where \(f\) denotes the activation function, and \(W\) and \(b\) represent the weights and biases at each layer, respectively.

Machine learning models in this context assume that historical data contain intrinsic patterns that can be leveraged to forecast future market behavior. Unlike traditional econometric models, machine learning approaches typically do not assume a specific functional form or distribution of errors but require careful tuning of hyperparameters and validation to prevent overfitting. The relationships among these machine learning models vary from the simplicity and interpretability of linear models to the complexity and predictive power of ensemble and neural network approaches. The trade-offs between bias and variance, interpretability and accuracy, and the computational cost of model training and prediction are central to the selection and application of these methodologies.

In the context of energy markets, where the influencing factors are numerous and the relationships between them can be highly nonlinear and interactive, the learning capability of machine learning models may be beneficial. Decision trees, ensemble methods, and neural networks, especially, stand out for their ability to model nonlinear relationships without extensive prespecification and for their robustness against outliers and variability in data quality.

\section{Empirical results}

\subsection{In-sample performance}

Table 2 provides the in-sample estimates for three univariate GARCH models discussed in section 3. First, beginning with the standard GARCH(1,1) model, it estimates the conditional variance as a function of past squared errors and its own lagged values. For crude oil, the model parameters \(\alpha\) and \(\beta\), with standard errors enclosed in parentheses, indicate significant persistence in volatility, as \(\alpha+\beta\) approaches unity. This indicates that shocks to the crude oil market have a lasting effect, recalling the findings of \citet{engle1982,bollerslev1986} regarding the long memory characteristic of time series volatility. This phenomenon can also be observed in the estimates of the other three energy commodities, thereby implying high degrees of volatility persistence in energy price processes across different energy commodities.

\begin{table}[ht]
\centering
\caption{Estimation results of different univariate GARCH models.}
\resizebox{\linewidth}{!}{\begin{tabular}{lllllllllllll}
\toprule
& \multicolumn{4}{l}{GARCH(1,1)} & \multicolumn{4}{l}{EGARCH} & \multicolumn{4}{l}{GJR-GARCH} \\
\hline
& CO & G & HO & NG & CO & G & HO & NG & CO & G & HO & NG \\ 
\midrule
$\mu$ & 0.000 & 0.001 & 0.000 & -0.001 & 0.000 & 0.001 & 0.000 & -0.001** & 0.000 & 0.001 & 0.000 & -0.001 \\
& (0.000) & (0.000) & (0.000) & (0.000) & (0.000) & (0.000) & (0.000) & (0.000) & (0.000) & (0.000) & (0.000) & (0.000) \\
$\omega$ & 0.000** & 0.000*** & 0.000 & 0.000*** & -0.084*** & -0.210*** & -0.043*** & -0.179*** & 0.000* & 0.000 & 0.000 & 0.000 \\
& (0.000) & (0.000) & (0.000) & (0.000) & (0.009) & (0.038) & (0.000) & (0.032) & (0.000) & (0.000) & (0.000) & (0.000)  \\
$\alpha$ & 0.068*** & 0.145*** & 0.049 & 0.167*** & -0.072*** & -0.004 & -0.039*** & -0.004 & 0.019*** & 0.144*** & 0.029*** & 0.163*** \\
& (0.009) & (0.014) & (0.038) & (0.014) & (0.018) & (0.011) & (0.007) & (0.011) & (0.007) & (0.017)& (0.010) & (0.017) \\
$\beta$ & 0.924*** & 0.831*** & 0.946*** & 0.824*** & 0.989*** & 0.967*** & 0.994*** & 0.970*** & 0.935*** & 0.831*** & 0.950*** & 0.824*** \\
& (0.010) & (0.015) & (0.042) & (0.013) & (0.001) & (0.006) & (0.000) & (0.005) & (0.010) & (0.015) & (0.011) & (0.013) \\
$\gamma$ & & & & & 0.109*** & 0.280*** & 0.088*** & 0.295*** & 0.076*** & 0.001 & 0.033*** & 0.009 \\
 & & & & & (0.012) & (0.021) & (0.003) & (0.017) & (0.012) & (0.020) & (0.009) & (0.021) \\
\bottomrule
$Q^2$(40) & 20.042 & 20.676 & 35.791*** & 45.916*** & 37.974*** & 24.505 & 55.285*** & 57.931*** & 28.319* & 20.676 & 39.410*** & 46.541*** \\
& [0.501] & [0.453] & [0.007] & [0.000] & [0.003] & [0.212] & [0.000] & [0.000] &[0.079] & [0.453] & [0.002] & [0.000] \\
ARCH & 18.651 & 18.849 & 23.603 & 35.790*** & 23.427 & 18.285 & 19.006 & 30.918** & 22.235 & 18.846 & 22.243 & 35.463*** \\
(40)& [0.603] & [0.588] & [0.255] & [0.007] & [0.264] & [0.631] & [0.575] & [0.035] &[0.337] & [0.588] & [0.337] & [0.007] \\
\hline
\vspace{-1cm}
\end{tabular}}
\caption*{Notes: Numbers in parentheses are standard errors and numbers in square brackets are p-values. $Q^2$ and ARCH refer to weighted Ljung-Box Q statistics of order 40 on the squared standard residuals and weighted ARCH-LM test statistics of order 40. *, **, *** denote significance at 10\%, 5\%, and 1\%, respectively.}
\end{table}

The EGARCH model, designed to capture the asymmetries and leverage effects in volatility (\citet{nelson1991}), delivers a significant positive coefficient \(\gamma\) for all four energy commodities, suggesting asymmetric effects of standardized shocks to volatilities, where negative shocks have a more pronounced impact on volatility than positive ones of the same magnitude. However, the GJR-GARCH model, similarly incorporating an additional term \(\gamma\) to account for the asymmetric effects of shocks (\citet{glosten1993}), reveals significantly positive \(\gamma\) values for crude oil and heating oil.

The results of the diagnostic tests on the squared residuals are presented in the lower part of Table 2. For crude oil and gasoline, \citet{ljung1978}'s \(Q^2(40)\) and \citet{engle1982}'s ARCH(40) test statistics cannot reject the null hypothesis of no serial correlation and homoscedastic error terms at the 10\% level, reflecting the models' ability to adequately capture the serial correlation in squared residuals and the presence of ARCH effects. However, for heating oil and natural gas, especially for natural gas, the results are consistent across the models, and none of the three models can fit the return series well.

For the GARCH(1,1) model, the sufficient condition for the existence of the second moment, and thus finite and well-defined variance, is that \(\alpha+\beta<1\) (\citet{ling2003}). Across all commodities, the sum of these parameters is below one, indicating that the second moment exists and the series are variance stationary under the GARCH(1,1) framework. In the EGARCH model, the values of \(|\beta|\) for each return series are less than one, supporting the existence of the second moment. For the GJR-GARCH model, similar to GARCH(1,1), considering \(\alpha + \beta + \gamma/2\), the combined value is expected to be less than one for variance stationarity. The reported values suggest that this condition is met, ensuring the presence of a second moment.

\renewcommand{\arraystretch}{1}
\begin{table}[ht]
\centering
\caption{Estimation results of different univariate GARCH-X models with external variables.}
\resizebox{\linewidth}{!}{\begin{tabular}{lllllllllllll}
\toprule
& \multicolumn{4}{l}{GARCH(1,1)-X} & \multicolumn{4}{l}{EGARCH-X} & \multicolumn{4}{l}{GJR-GARCH-X} \\
\hline
& CO & G & HO & NG & CO & G & HO & NG & CO & G & HO & NG \\ 
\midrule
$\mu$ & 0.000*** & 0.000 & 0.001*** & 0.000*** & 0.000 & 0.001 & 0.000 & -0.001** & 0.000 & 0.001* & 0.000*** & -0.000 \\
$\omega$ & 0.000 & 0.000*** & 0.000*** & 0.000*** & -0.096*** & -0.209*** & -0.073*** & -0.222*** & 0.000*** & 0.000*** & 0.000*** & 0.000*** \\
$\alpha$ & 0.050*** & 0.156*** & 0.123*** & 0.126*** & -0.059*** & 0.001 & -0.029*** & -0.004 & 0.085*** & 0.189*** & 0.166*** & 0.158*** \\
$\beta$ & 0.900*** & 0.689*** & 0.339*** & 0.852*** & 0.988*** & 0.967*** & 0.991*** & 0.965*** & 0.023** & 0.633*** & 0.309*** & 0.794*** \\
$\gamma$ & & & & & 0.074*** & 0.257*** & 0.085*** & 0.282*** & 0.041* & -0.021 & -0.087*** & -0.001 \\
$\sigma_{Co}$ & & 0.000 & 0.426*** & -0.020***&  & 0.999 & 0.999 & 0.999 &  & 0.268*** & 0.474*** & -0.011\\
$\sigma_{G}$ & 0.001*** & & 0.025***& 0.007***& 0.391 & & 0.999 & 0.999 & 0.016*** &  & 0.018*** & 0.006* \\
$\sigma_{HO}$ & 0.042***& 0.108*** & & 0.056***& 0.999 & 0.999 &  & 0.999 & 0.886*** & 0.103*** &  & 0.057** \\
$\sigma_{NG}$ & 0.000***& -0.002*** & -0.000*** & & -0.028 & -0.999 & -0.778* &  & -0.001*** & -0.002*** & 0.005*** &  \\
$INV_{CO}$ & 0.000***& -0.002 & 0.003***& 0.004***& -0.999 & 0.999 & 0.999 & -0.999 & 0.010*** & 0.005 & -0.004*** & -0.004*** \\
$INV_{G}$ & -0.000***& 0.001 & 0.001***& 0.004***& -0.529 & 0.566 & -0.158& 0.999 & -0.001*** & 0.002** & 0.001*** & 0.003*** \\
$INV_{HO}$ & 0.000***& -0.005*** & -0.000***& 0.001***& -0.626 & -0.999 & -0.999 & 0.999 & 0.001*** & -0.004** & -0.000*** & 0.001 \\
$INV_{NG}$ & 0.000***& -0.002*** & -0.001***& -0.001***& 0.235 & -0.776* & -0.063 & -0.999** & -0.001*** & -0.002*** & -0.001*** & -0.002*** \\
DI & 0.003***& 0.005* & -0.001***& 0.010***& 0.999 & 0.999 & 0.999 & -0.999 & -0.000 & 0.011*** & -0.001*** & 0.005*** \\
EFFR & -0.000***& 0.000 & 0.000***& -0.002***& -0.492*** & -0.228 & 0.000 & -0.210 & 0.000*** & -0.000 & 0.000*** & -0.000 \\
SP500 & -0.000***& 0.004*** & 0.000***& 0.007***& -0.999 & -0.999 & -0.999 & 0.999 & -0.001*** & 0.003** & 0.001*** & 0.004*** \\
T10Y3M & -0.000***& -0.000** & 0.000***& 0.000***& -0.293** & 0.325* & -0.208 & -0.193 & 0.000*** & 0.000* & -0.000*** & 0.000 \\
CPI & 0.004***& 0.095*** & 0.077***& 0.062***& 0.999 & -0.999 & -0.999 & -0.999 & 0.054*** & 0.132*** & 0.134*** & 0.002 \\
IP & 0.001***& 0.015*** &0.006*** & -0.009***& -0.999 & -0.999 & 0.999 & 0.999 & -0.011* & 0.025 & 0.011*** & 0.002 \\
PPI & -0.001***& -0.058*** & -0.001***& -0.024***& -0.798*** & -0.999 & -0.999 & -0.999 & -0.028***& -0.062*** & -0.019*** & -0.017 \\
UNRATE & -0.000***& -0.000** &0.000*** & 0.000***& 0.489** & 0.309 & 0.429* & 0.534 & 0.001*** & 0.000 & -0.000*** & 0.000*** \\
GEPU & 0.000***& -0.000 & 0.000***& 0.001***& 0.741*** & 0.468 & 0.240 & 0.715* & -0.000 & -0.000 & 0.001*** & 0.001 \\
TEMP & 0.000***& 0.000 & -0.001***& -0.001***& 0.299 & 0.152 & 0.050 & -0.312 & -0.000 & 0.001 & -0.001*** & -0.001 \\
\hline
\vspace{-1cm}
\end{tabular}}
\caption*{Notes: *, **, *** denote significance at 10\%, 5\%, and 1\%, respectively.}
\end{table}

Table 3 presents in-sample estimates for three univariate GARCH models with external variables, as discussed in Section 3. These models extend the standard GARCH(1,1) model by incorporating external variables to better understand how exogenous factors influence the conditional variance of energy commodities.

Starting with the GARCH(1,1)-X model, the inclusion of exogenous variables appears to significantly affect the volatility estimates, as indicated by the significant coefficients for the external variables across different commodities. This implies that factors beyond past squared errors and lagged values are influencing the conditional variance, aligning with broader economic theories that external factors such as market indices and macroeconomic indicators impact energy prices. For instance, the significant positive coefficient for the S\&P 500 index for gasoline, heating oil, and natural gas suggests that rising equity markets, indicative of economic optimism, can lead to increased volatility in these commodities, possibly due to anticipatory behaviors among traders. The negative coefficients for the federal funds rate across most commodities suggest that tighter monetary policy, indicated by higher interest rates, is associated with lower commodity volatility, possibly due to dampened economic activity and reduced speculative trading. The global economic policy uncertainty index (GEPU) and temperature anomalies (TEMP) also exhibit significant coefficients, highlighting the broader economic and environmental factors affecting the volatility of energy commodity markets. The significance of these external variables varies across commodities, suggesting differentiated impacts on their volatility. The results are similar to the GJR-GARCH-X model. The EGARCH-X model, on the other hand, seemed to weigh external variables differently, rendering most of them insignificant.

\begin{table}[ht]
\centering
\caption{Estimation results of multivariate GARCH-BEKK model}
\resizebox{\textwidth}{!}{\begin{tabular}{lll}
\hline
\multicolumn{3}{l}{Panel A: Conditinal variance-covariance structure} \\ 
\hline \\[0.1pt]
\vspace{0.7cm}
C=$\begin{bmatrix} 0.474^{***} &  &  &  \\ 
                    (0.015) & & & \\
                    0.176^{***} & 0.499^{***} &  &  \\ 
                    (0.032) & (0.020) & & \\
                    0.149^{***} & 0.023 & 0.079^{***} &  \\ 
                    (0.015) & (0.016)& (0.019) & \\
                    -0.052 & -0.001 & -0.166 & 0.708^{***} \\ 
                    (0.048) & (0.059) & (0.160) & (0.050) \end{bmatrix}$ 
& A=$\begin{bmatrix} 0.387^{***} & -0.035^{**} & 0.006 & -0.038^{***} \\ 
                    (0.009) & (0.016) & (0.006)& (0.014)\\
                    -0.037^{***} & 0.329^{***} & -0.020^{***}&  -0.056^{***}\\ 
                    (0.006) & (0.008) & (0.005)& (0.008)\\
                    -0.100^{***} & -0.075^{***} & 0.207^{***} & 0.043^{**} \\ 
                    (0.010) & (0.012)& (0.006) & (0.018)\\
                    -0.001 & 0.005 & 0.002 & 0.398^{***} \\ 
                    (0.003) & (0.005) & (0.003) & (0.006) \end{bmatrix}$ 
& G=$\begin{bmatrix} 0.895^{***} & 0.011^{*} & -0.003 & 0.019^{***} \\ 
                    (0.003) & (0.006) & (0.003)& (0.006)\\
                    0.010^{***} & 0.939^{***} & 0.005^{***}&  0.014^{***}\\ 
                    (0.002) & (0.003) & (0.002)& (0.003)\\
                    0.045^{***} & 0.021^{***} & 0.977^{***} & -0.020^{***} \\ 
                    (0.003) & (0.004)& (0.002) & (0.006)\\
                    0.001 & -0.002 & 0.001 & 0.916^{***} \\ 
                    (0.002) & (0.002) & (0.001) & (0.003) \end{bmatrix}$ \\
\hline
\end{tabular}}
\caption*{Notes: Numbers in parentheses are standard errors. Multivariate Ljung-Box $Q^2$(40) statistics of order 40 on the squared standard residuals amount to 4.123 with a $p$-value of $0.999$. \\
*, **, *** denote significance at 10\%, 5\%, and 1\%, respectively.}
\end{table}

The estimation results of the multivariate GARCH-BEKK model presented in Table 4 shed light on the dynamic interrelations and volatility spillovers among four energy commodities. The BEKK model's conditional variance-covariance structure is incorporated into the matrices \(C\), \(A\), and \(G\), each offering distinct insights into market dynamics. The matrix \(C\), representing the constant term in the variance-covariance equation, contains significant diagonal elements. These values indicate baseline volatilities that are inherent to each commodity, reflecting market-specific risks and uncertainties.

The matrix \(A\) details the response of current volatility to past shocks. The significant diagonal elements represent the persistence of shocks within each commodity market. Notably, there are statistically significant volatility spillover effects in some of the off-diagonal elements. For instance, there is volatility spillover from crude oil to the gasoline and natural gas markets. Gasoline and heating oil also exhibit volatility spillover to natural gas. However, the volatility in the natural gas market doesn't seem to have any impact on the other three commodity markets.

Matrix \(G\) represents the autoregressive component of conditional variances, emphasizing the persistence of volatility over time. The high diagonal values indicate a strong autoregressive effect, suggesting that volatilities in these markets are highly persistent. The off-diagonal elements, although smaller, denote the extent of volatility spillovers across markets, with positive values indicating the transmission of volatility from one commodity to another. Here again, there are volatility spillover effects from the petroleum products to natural gas, but not the other way around.

\subsection{Out-of-sample forecast performance}

The evidence from within the sample provides a historical performance of models. Nevertheless, the performance outside the sample is considered more critical since market participants are primarily interested in the prospective efficiency of the models. Hence, we explore the out-of-sample forecasting capabilities of these models. The procedure for forecasting is outlined as follows:

Our data spans from January 23, 2006, to December 29, 2023, encompassing a total of 4,506 data points. Of these, 1,000 data points, starting from January 2, 2020, are reserved for out-of-sample analysis. The remaining 3,506 data points constitute the in-sample, used for the initial training and validation, with a 20\% portion of the training set split for validation in machine learning models. In contrast, for GARCH models, all 3,506 in-sample data points are utilized for training. The estimation process also adopts a rolling window approach, recalculating the model parameters daily to predict the next day's volatility.

As outlined by \citet{bollerslev1994, diebold1996, lopez2001}, it is not clear which loss function best assesses the accuracy of volatility model forecasts. Considering this, instead of choosing just one metric, I include both root mean squared error (RMSE) and mean absolute error (MAE). RMSE is a standard measure used to evaluate the differences between values predicted by a model and the values observed. It is defined as:

\[ RMSE = \sqrt{\frac{1}{n}\sum_{t=1}^{n}(\sigma_t - \hat{\sigma}_t)^2} ,\]

\noindent where \(\sigma_t\) represents the actual observed values, \(\hat{\sigma}_t\) denotes the forecasted values, and \(n\) is the number of forecasts. RMSE values are always non-negative, and a value of 0 would indicate a perfect fit. In energy market volatility forecasting, a lower RMSE value indicates the model's higher accuracy.

MAE measures the average magnitude of errors in a set of predictions, without considering their direction:

\[ MAE = \frac{1}{n}\sum_{t=1}^{n} |\sigma_t - \hat{\sigma}_t| .\]

Furthermore, according to \citet{nomikos2011}, these standard statistics do not differentiate between the impacts of over-forecasting versus under-forecasting volatility, which can significantly affect market decisions, especially in scenarios like crude oil pricing and its effects on call option values. Hence, I add mean mixed error overprediction (MMEO) and mean mixed error underprediction (MMEU) following \citet{brailsford1996}:

MMEO and MMEU are metrics designed to evaluate the bias in forecasts, specifically the tendency of a model to overpredict or underpredict. They can be defined as:

\[ MMEO = \frac{1}{n}\sum_{t=1}^{n} (\hat{\sigma}_t > \sigma_t) * (\hat{\sigma}_t - \sigma_t) \]
\[ MMEU = \frac{1}{n}\sum_{t=1}^{n} (\hat{\sigma}_t < \sigma_t) * (\sigma_t - \hat{\sigma}_t) ,\]

\noindent where the terms \((\hat{\sigma}_t > \sigma_t)\) and \((\hat{\sigma}_t < \sigma_t)\) operate as indicators, becoming 1 if the condition is true and 0 otherwise. MMEO assesses the degree to which a model's predictions exceed the actual values, while MMEU measures the extent to which predictions fall short.

\begin{table}[ht]
\centering
\caption{Out-of-sample forecasting performance of GARCH and machine learning models.}
\resizebox{\linewidth}{!}{\begin{tabular}{@{}lllllllllllllllll@{}} 
\toprule
\multirow{2}{*}{GARCH} & \multicolumn{4}{l}{CO} & \multicolumn{4}{l}{G} & \multicolumn{4}{l}{HO} & \multicolumn{4}{l}{NG} \\
\cline{2-17}
 & RMSE & MAE & MMEO & MMEU & RMSE & MAE & MMEO & MMEU & RMSE & MAE & MMEO & MMEU & RMSE & MAE & MMEO & MMEU \\ 
\midrule
GARCH(1,1) & 0.230 & 0.052 & 0.093 & 0.019 & 0.044 & 0.037 & 0.038 & 0.002 & 0.032 & 0.029 & 0.030 & 0.001 & 0.077 & 0.060 & 0.063 & 0.008 \\
EGARCH & 0.221 & 0.045 & 0.086 & 0.016 & 0.041 & 0.036 & 0.037 & 0.002 & 0.030 & 0.028 & 0.028 & 0.001 & 0.068 & 0.055 & 0.058 & 0.007 \\
GJR-GARCH & 0.232 & 0.055 & 0.097 & 0.020 & 0.044 & 0.037 & 0.039 & 0.002 & 0.032 & 0.029 & 0.030 & 0.001 & 0.078 & 0.060 & 0.063 & 0.008\\
GARCH-X & 0.223 & 0.038 & 0.079 & 0.015 & 0.033 & \thicker{0.027} & 0.028 & 0.002 & 0.014 & 0.011 & 0.011 & 0.000 & 0.075 & 0.057 & 0.061 & 0.008 \\
EGARCH-X & 0.219 & 0.040 & 0.082 & 0.013 & 0.040 & 0.036 & 0.037 & 0.002 & 0.028 & 0.026 & 0.027 & 0.001 & \thicker{0.065} & \thicker{0.053} & 0.056 & 0.007 \\
GJR-GARCH-X & \thicker{0.211} & \thicker{0.021} & 0.062 & 0.010 & \thicker{0.033} & 0.027 & 0.028 & 0.002 & \thicker{0.013} & \thicker{0.010} & 0.011 & 0.000 & 0.070 & 0.054 & 0.057 & 0.007 \\
GARCH-BEKK & 0.227 & 0.063 & 0.105 & 0.016 & 0.055 & 0.050 & 0.051 & 0.004 & 0.039 & 0.037 & 0.038 & 0.002 & 0.075 & 0.066 & 0.069 & 0.008 \\
\hline
\multirow{2}{*}{Machine Learning} & \multicolumn{4}{l}{CO} & \multicolumn{4}{l}{G} & \multicolumn{4}{l}{HO} & \multicolumn{4}{l}{NG} \\
\cline{2-17}
 & RMSE & MAE & MMEO & MMEU & RMSE & MAE & MMEO & MMEU & RMSE & MAE & MMEO & MMEU & RMSE & MAE & MMEO & MMEU \\ 
\midrule
Linear Regression & \thicker{0.219} & 0.015 & 0.016 & 0.052 & 0.155 & 0.017 & 0.017 & 0.024 & 0.020 & 0.003 & 0.003 & 0.000 & 0.050 & 0.011 & 0.013 & 0.003 \\
Ridge & 0.222 & 0.011 & 0.011 & 0.049 & 0.016 & 0.004 & 0.004 & 0.000 & 0.004& 0.001& 0.001& 0.000 & \thicker{0.039} & 0.008 & 0.010 & 0.002 \\
Lasso & 0.222 & 0.011 & 0.011 & 0.049 & 0.010 & 0.002 & 0.003 & 0.000 & 0.004 & 0.001 & 0.001 & 0.000 & 0.044 & 0.007 & 0.007 & 0.002 \\
Elastic Net & 0.222 & 0.011 & 0.011 & 0.049 & 0.010 & 0.002 & 0.003 & 0.000 & 0.004 & 0.001 & 0.001 & 0.000 & 0.044 & 0.007 & 0.007 & 0.002 \\
Bayesian Ridge & 0.220 & 0.013 & 0.013 & 0.050 & 0.010 & 0.002 & 0.003 & 0.000 & 0.004 & 0.001 & 0.001 & 0.000
 & 0.039 & 0.008 & 0.010 & 0.002 \\
Decision Tree & 0.222 & 0.011 & 0.011 & 0.049 & 0.010 & 0.002 & 0.003 & 0.000 & 0.004 & 0.001 & 0.001 & 0.000
 & 0.043 & 0.007 & 0.008 & 0.002 \\
Random Forest & 0.222 & 0.011 & 0.011 & 0.049 & 0.012 & 0.003 & 0.003 & 0.000 & 0.003 & \thicker{0.001} & 0.002 & 0.000 & 0.041 & 0.008 & 0.009 & 0.002 \\
XGBoost & 0.222 & \thicker{0.010} & 0.011 & 0.049 & \thicker{0.010} & \thicker{0.002} & 0.003 & 0.000 & \thicker{0.003} & 0.001 & 0.001 & 0.000
 & 0.040 & \thicker{0.007} & 0.008 & 0.002 \\
SVM & 0.222 & 0.020 & 0.021 & 0.049 & 0.011 & 0.007 & 0.008 & 0.000 & 0.010 & 0.009 & 0.010 & 0.000
 & 0.043 & 0.010 & 0.012 & 0.002 \\
KNN & 0.222 & 0.011 & 0.011 & 0.049 & 0.010 & 0.002 & 0.003 & 0.000 & 0.004 & 0.001 & 0.001 & 0.000
 & 0.043 & 0.007 & 0.007 & 0.002 \\
MLP & 0.222 & 0.011 & 0.012 & 0.049 & 0.011 & 0.004 & 0.005 & 0.000 & 0.004 & 0.001 & 0.001 & 0.000
 & 0.040 & 0.009 & 0.009 & 0.002 \\
\hline
\end{tabular}}
\end{table}

Table 5 reports the out-of-sample forecasting performance of traditional GARCH models and various machine learning models across four energy commodities. GARCH models, including GARCH(1,1), EGARCH, GJR-GARCH, and their extensions with external variables (denoted as GARCH-X), exhibit varying degrees of forecasting performance. For instance, the GJR-GARCH model demonstrates relatively higher RMSE and MAE values across all commodities, with RMSE values of 0.232, 0.044, 0.032, and 0.078 for crude oil, gasoline, heating oil, and natural gas, respectively. This suggests a slightly less accurate forecast compared to other GARCH models in the context of this analysis. On the contrary, the GJR-GARCH-X model, incorporating external variables, improves significantly, evidenced by a notable reduction in RMSE for natural gas to 0.070, indicating the beneficial impact of including additional explanatory factors in volatility forecasting. Furthermore, univariate GARCH variants without external regressors, especially EGARCH, largely outperform the multivariate GARCH-BEKK model in terms of all metrics considered. This may be due to the overparametrization of the model with four interacting energy market variables, which may lead to parameter instability when estimated. Machine learning models, on the other hand, show similar performance to each other. Except for linear regression, which is undoubtedly the worst model among the machine learning models, others show similar values of metrics across all four energy commodities.

Except for crude oil, machine learning models consistently outperform their GARCH counterparts, including the multivariate GARCH-BEKK model, in terms of RMSE and MAE. With respect to MMEO and MMEU, there is a clear trend that machine learning models tend to underpredict, while GARCH models tend to overpredict. The difference becomes more evident when crude oil is considered. In fact, crude oil is the only energy commodity that machine learning models fail to beat GARCH variants in terms of RMSE. The GJR-GARCH-X model reports the lowest RMSE value of 0.211. A possible explanation for this could be that the crude oil spot price experienced a record negative spot price on April 20, 2020, such that this huge volatility caused a large error when squared. However, GARCH variants tend to overpredict and, as Figure 3 illustrates, the GJR-GARCH-X model manages to reduce the gap between the actual huge volatility value and the predicted value by making a jump on April 20, 2020. Apart from that, GARCH models show weak performance across the energy commodities. For instance, the GJR-GARCH model's MAE for crude oil is notably high at 0.055, indicating a larger average error in forecasts compared to 0.011 of the best-performing machine learning models. Similar trends are observed with other GARCH variants like EGARCH and GJR-GARCH, where, despite adjustments for asymmetric information and external regressors, the error metrics, while competitive, do not consistently outperform their machine learning counterparts.

The consistent outperformance of machine learning models across the performance evaluation metrics for various energy commodities not only attests to their superior forecasting capability but also implies the potential limitations of traditional GARCH models in adapting to the complex, non-linear patterns observed in energy market volatility. The machine learning models' lower error metrics reflect their robustness in data processing and pattern recognition in markets characterized by rapid changes and high uncertainty.

\begin{figure}[!ht]
\centering
\caption{Out-of-sample volatility forecasts based on GJR-GARCH-X and XGBoost model}
\includegraphics[width=\textwidth, height=.85\textheight]{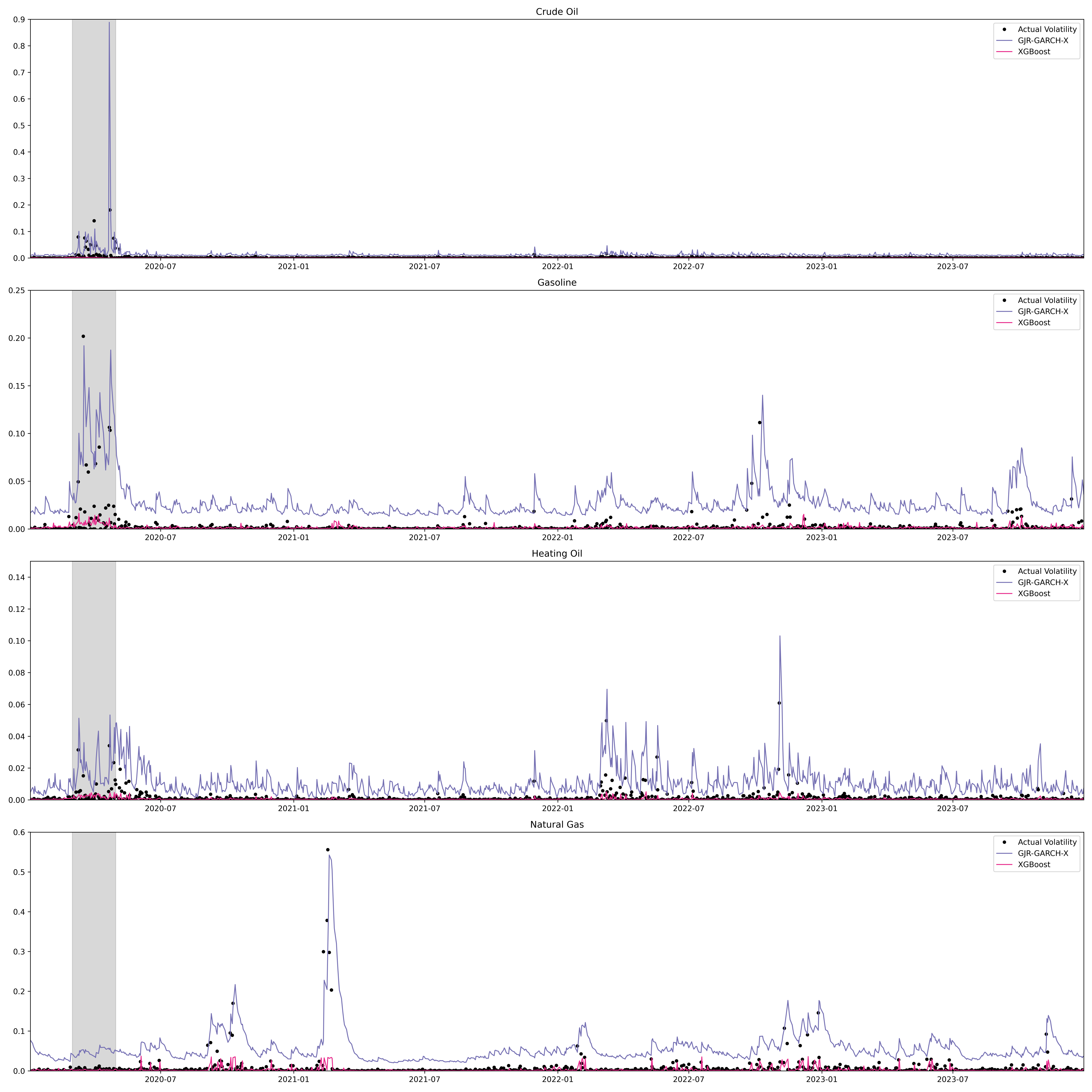}\\
\caption*{The figure illustrates out-of-sample volatility forecasts of crude oil, gasoline, heating oil, and natural gas for GJR-GARCH-X and XGBoost models. The grey shaded areas depict NBER recession months.}
\end{figure}

Figure 3 illustrates the out-of-sample volatility forecasts generated by the GJR-GARCH model and the XGBoost model as the best performing model for each group. The GJR-GARCH model's strength lies in its structural foundation, which explicitly accounts for the leverage effect—a phenomenon where negative shocks to returns have a more pronounced impact on future volatility than positive shocks of equivalent magnitude. This attribute is particularly evident in the figure, where the GJR-GARCH model closely follows the trajectory of jumps in actual volatility, albeit with limitations in capturing the full magnitude of the spikes.

In contrast, the forecasts of the XGBoost model, while generally smoother and less reactive to individual spikes than the GJR-GARCH model, exhibit a higher level of adherence to the underlying volatility trend. This observation suggests that machine learning models, through their inherent flexibility and ability to discern complex patterns from vast datasets, can effectively encapsulate the underlying dynamics of energy market volatility, albeit with a tendency toward underestimating the extremities of volatility spikes.

\subsection{Variable importance}

The Shapley Additive Explanations (SHAP) values, rooted in cooperative game theory, offer a consistent method to quantify the contribution of each feature to the prediction of a complex machine learning model. Developed from the Shapley value framework proposed by \citet{shapley1953}, SHAP values offer an equitable solution to distributing the "payout" – the prediction output – among the "players" – the input variables.

The Shapley value for each feature is calculated based on its marginal contribution across all possible combinations or coalitions of features. This approach ensures that the contributions are distributed fairly, adhering to principles like efficiency, symmetry, dummy, and additivity, which form the core of the Shapley value theory. Specifically, for machine learning interpretations, the contribution of a feature value to the predictive model is assessed by computing the average marginal contribution across all possible subsets:

\[ \phi_j(v) = \sum_{S \subseteq N \setminus \{j\}} \frac{|S|!(|N|-|S|-1)!}{|N|!} (v(S \cup \{j\}) - v(S)) \]

Here, \(\phi_j(v)\) represents the SHAP value for feature \(j\), indicating its impact on the model’s output. \(N\) is the total set of features, and \(S\) represents a subset of these features excluding feature \(j\). The term \(v(S)\) denotes the prediction for the subset \(S\), thereby illustrating the incremental value added by including feature \(j\) in the prediction process.

I focus on the Shapley values of the XGBoost model as the best performing model, on average, approximated by the TreeSHAP algorithm. TreeSHAP is an algorithm within the SHAP package in Python designed to explain predictions of tree-based machine learning models by efficiently calculating Shapley values. Shapley values distribute a model's output among its input features, showing each feature's contribution to the prediction. TreeSHAP optimizes this calculation by utilizing the tree structure, significantly speeding up the process compared to traditional methods. It does so by traversing the decision tree, calculating conditional expectations at each node, and then aggregating these to find the exact contribution of each feature. This approach not only enhances computational efficiency but also provides clear, actionable insights into how features influence model predictions, making complex models more interpretable.

Unlike KernelSHAP's model-agnostic approach, TreeSHAP leverages the decision paths within trees to directly compute SHAP values, eliminating the need for approximate sampling. For a given tree, the contribution of a feature in a single decision path is:

\[ \Delta \phi_j = \sum_{t \in T} \frac{(m_t - 1)! (M - m_t)!}{M!} \Delta p_t(x) (v_{tj}(x) - v_{t0}) ,\]

\noindent where \(T\) is the set of all possible paths in the tree, \(m_t\) is the number of features along path \(t\), \(M\) is the total number of features, \(\Delta p_t(x)\) is the change in prediction probability along path \(t\), \(v_{tj}(x)\) is the predicted value when feature \(j\) is included in the path, and \(v_{t0}\) is the predicted value without any features. By aggregating these values across all paths, TreeSHAP accurately assigns each feature its corresponding impact on the prediction.

\begin{figure}[!ht]
\caption{SHAP values based on XGBoost model for four different energy commodities}
    \centering
    \begin{tabular}{l l}
        (A) & (B) \\
        \includegraphics[width=.45\textwidth]{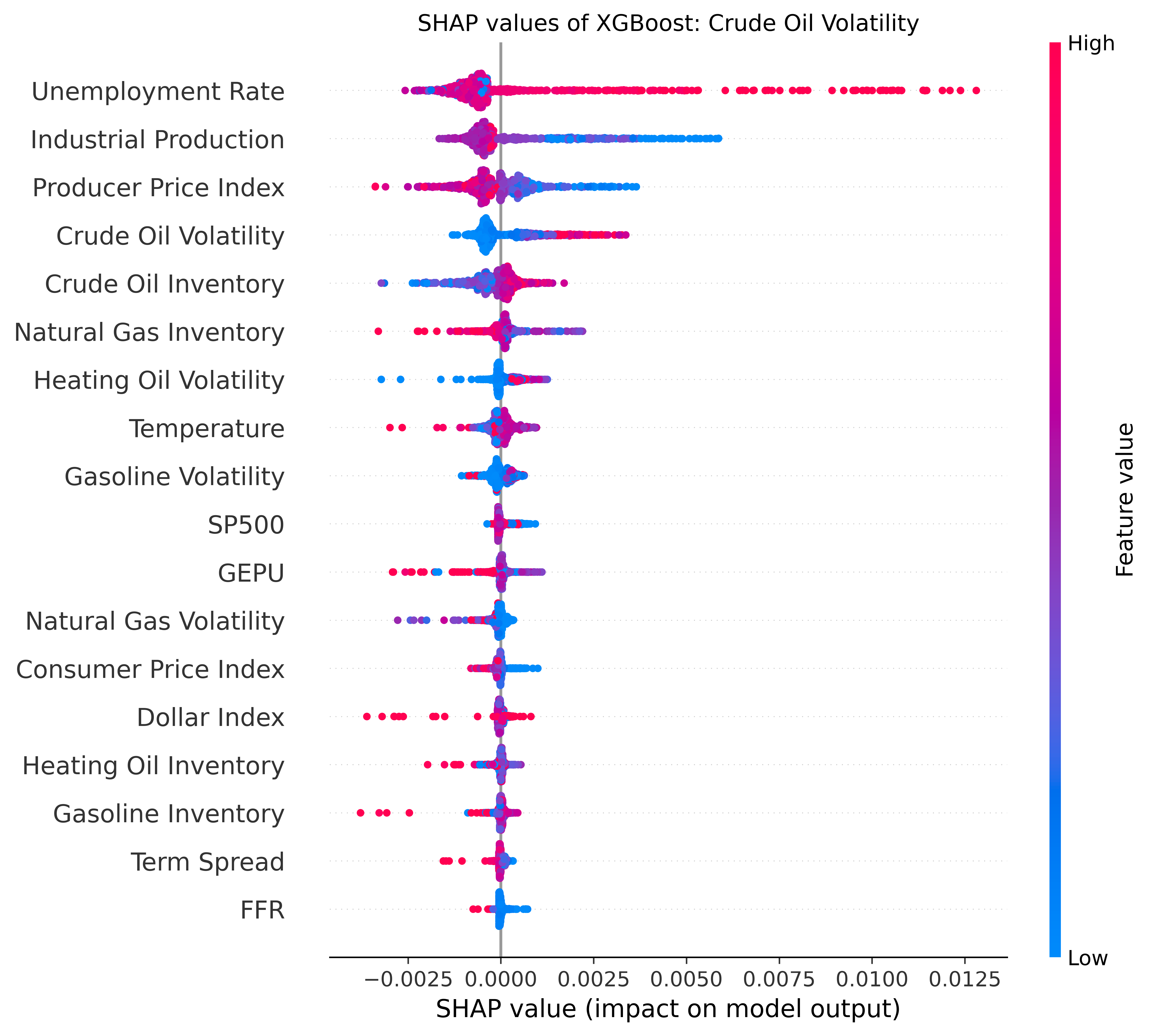} &  \includegraphics[width=.45\textwidth]{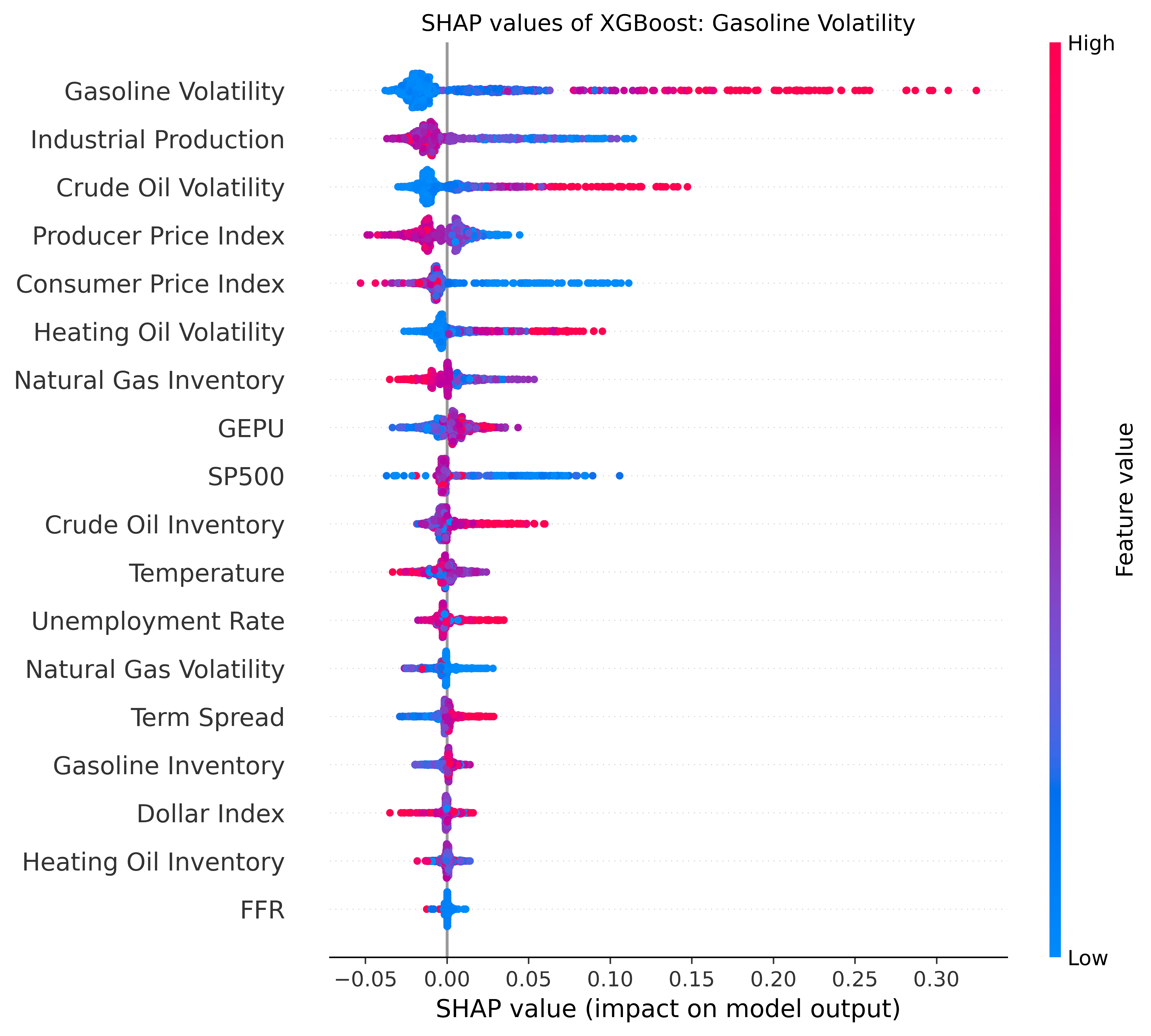} \\ 
         (D) & (E) \\
        \includegraphics[width=.45\textwidth]{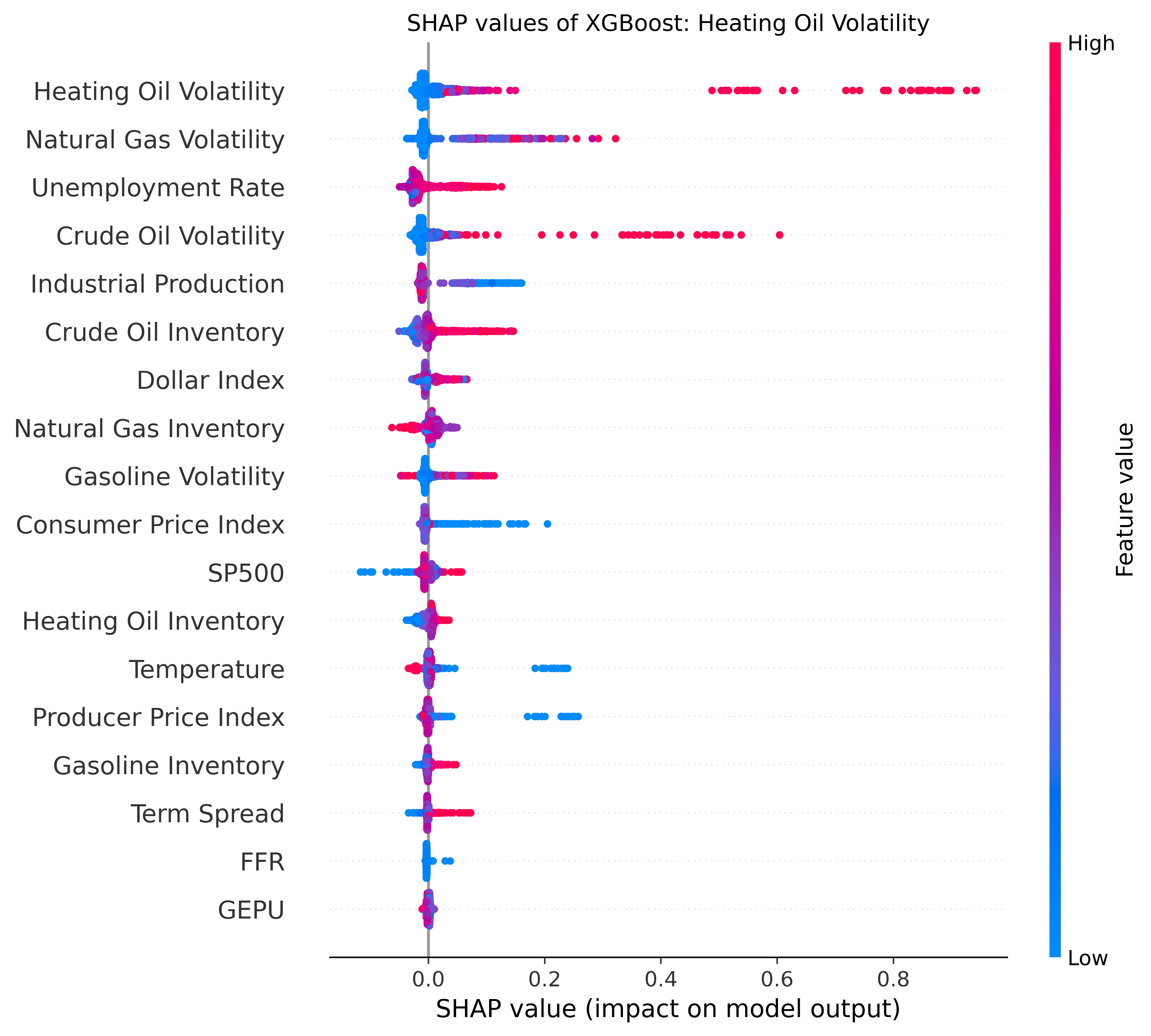} & \includegraphics[width=.45\textwidth]{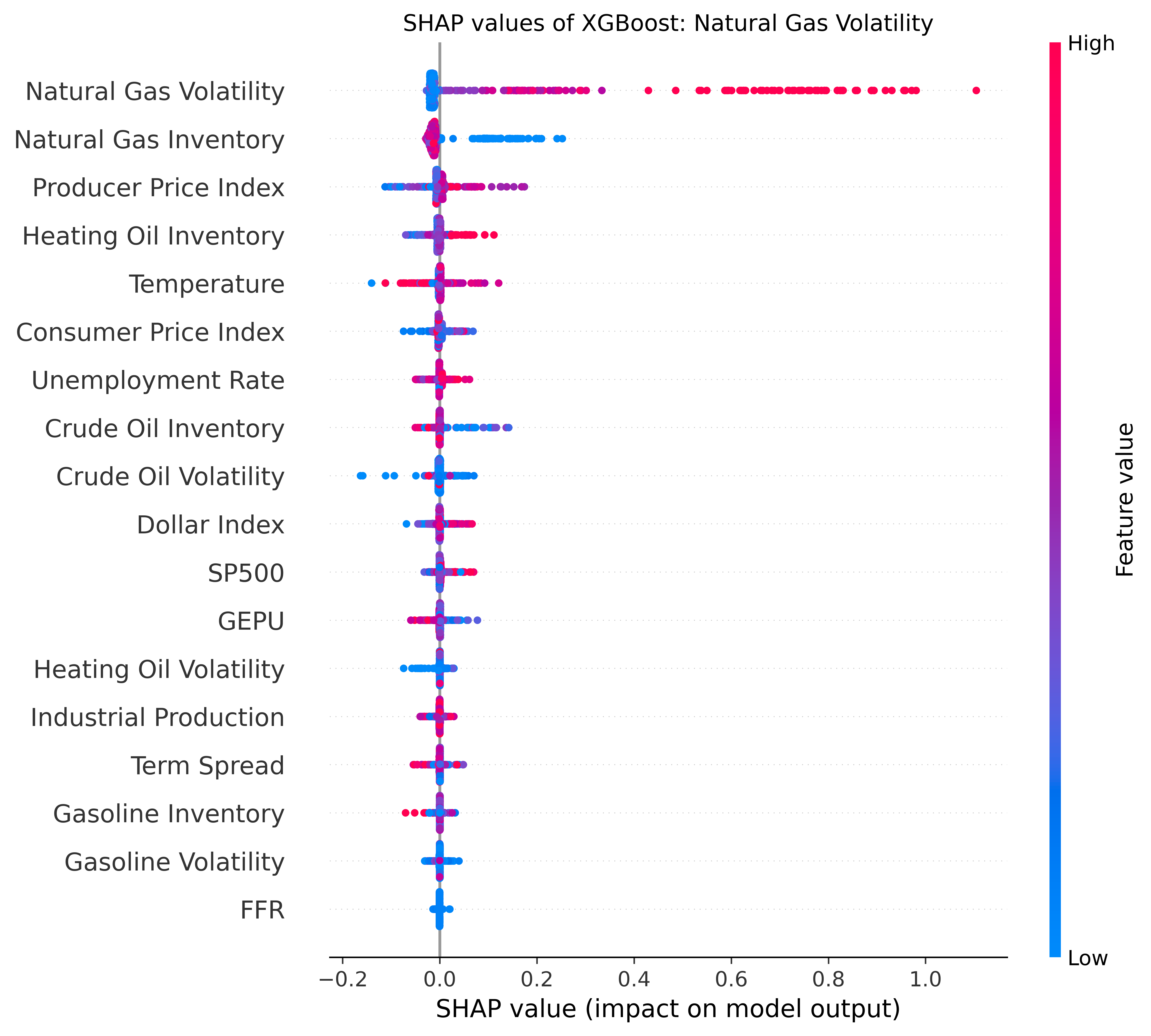} \\
    \end{tabular}
    \caption*{Panel (A) to (D) display XGBoost-based SHAP values of the predictors for crude oil, gasoline, heating oil, and natural gas volatility, respectively.}
\end{figure}

Figure 4 offers an in-depth look at the predictors' impact on model output for the XGBoost model, enabling a clearer understanding of model behavior. The four beeswarm plots provided for crude oil, gasoline, heating oil, and natural gas volatility offer a visual representation of how different variable values influence the prediction of volatility in these markets. Variables are ordered by the sum of the absolute SHAP values over all samples, with the most important variable at the top of the plot.

Gasoline volatility predictions seem to be highly dependent on their previous volatility values. High gasoline volatility value from the previous time step strongly positively affects the volatility prediction, which amounts to persistent volatilities and volatility clustering as illustrated in Figure 3, especially in the last quarter of 2023. There are also volatility spillover effects as depicted by crude oil and heating oil volatility affecting gasoline volatility positively. These results are in line with the in-sample estimation results for the GARCH-BEKK model in Table 4. This indicates that machine learning models can effectively capture both the persistent volatility structure and volatility transmission across different energy commodities, which has traditionally been examined only by multivariate GARCH models, and thereby offer significantly higher forecasting performance. From external regressors, the industrial production index and both producer and consumer price indexes, among others, seem to have strong explanatory power for the prediction output of gasoline volatility. However, the directions of effects of these variables differ from their signs of estimated coefficients from Table 3. Figure 4 indicates that all three variables affect gasoline volatility negatively, whereas Table 3 reports positive significant coefficients for consumer price and industrial production index. This can refer to different weight assignments between different types of models. Machine learning models can capture potential asymmetry and nonlinearity in economic data, which leads to different weight assignments.

The result is similar for heating oil volatility. It also exhibits a strongly persistent volatility structure, while there are volatility spillover effects from natural gas and crude oil markets. There is a clear signal that high previous crude oil volatility value positively affects the prediction output of heating oil volatility, whereas it is somewhat mixed for the volatility transmission of natural gas volatility. The unemployment rate and industrial production index have a positive effect on heating oil volatility.

Natural gas volatility is also highly persistent. However, the volatility spillover effects from other energy commodities are virtually nonexistent. In fact, the natural gas market seems to be isolated as the only volatility transmission captured by the XGBoost model is the volatility spillover from natural gas to heating oil market, but not the other way around. Both natural gas and heating oil are primary sources for heating purposes in residential and commercial buildings, especially in regions with cold climates. When the price or volatility of natural gas changes significantly, it can affect the demand and subsequently the price and volatility of heating oil as consumers and businesses may switch between these energy sources based on price, availability, and efficiency. This substitution effect can lead to volatility spillovers between these markets. However, apart from that, natural gas volatility is only influenced by its past values and inventory. The petroleum and natural gas markets possess distinct characteristics that can limit direct volatility transmission between them, due to differences in their primary uses, supply chain logistics, market participants, and geopolitical influences. Petroleum products are mainly utilized in transportation and manufacturing, while natural gas is predominantly used for heating and electricity generation. The infrastructure required for their production and distribution differs significantly, affecting how supply disruptions influence market volatility. Additionally, geopolitical sensitivities and regulatory policies can impact these markets in diverse ways, further distinguishing their volatility patterns. However, broader economic trends and major global events can still induce correlated movements across these energy markets.

Crude oil is the only energy commodity that shows particularly high correlations with macroeconomic variables. Crude oil volatility is less persistent compared to other energy commodities. Instead, a high unemployment rate and a low industrial production index, which may reflect poor economic conditions, increase crude oil volatility, while lower values of the producer price index also increase crude oil volatility. A low producer price index may imply high crude oil volatility by signaling an economic slowdown and reduced demand for oil, affecting investor sentiment and leading to speculative trading.

\section{Conclusion}

In this paper, energy price volatilities of crude oil, gasoline, heating oil, and natural gas are modeled and predicted using univariate and multivariate GARCH-class models, as well as machine learning models. The in-sample evidence of univariate GARCH models indicates that energy price volatility can be characterized by significant persistence and asymmetric effects. The in-sample estimation results of the multivariate GARCH-BEKK model indicate the persistence of shocks and volatility over time within each commodity market and volatility transmission across the different energy commodity markets. The relationship between crude oil and refining product markets can be characterized by volatility transmission, as there is volatility spillover from crude oil to gasoline and heating oil markets. The volatility transmission between the petroleum and natural gas market is less prevalent, as evidenced by estimates of the GARCH-BEKK model and SHAP values of the XGBoost model.

Machine learning models generally outperform traditional GARCH models in out-of-sample forecasting. This performance may be attributed to their capacity to capture nonlinear relationships. Particularly, Ridge, Lasso, XGBoost, and Random Forest models emerge as robust tools for forecasting energy market volatility. GARCH-class models tend to overpredict, whereas machine learning models tend to underestimate, thus GARCH-class models have better predictability for volatility spikes. It would be advisable to combine these two types of models and refer mostly to the predictions of the machine learning models, while taking predictions of spikes from GARCH-class models seriously.

The SHAP method delivers approximated Shapley values of the predictors for XGBoost as the best-performing model. It depicts the volatility transmission from crude oil to gasoline and heating oil markets. The crude oil market is less characterized by volatility persistence and transmission, instead showing greater dependence on economic conditions captured by the unemployment rate, industrial production index, and producer price index. Gasoline and heating oil volatility is highly persistent. There is volatility spillover from natural gas to the heating oil market. Apart from this connection, the natural gas market seems to be isolated, as natural gas volatility is heavily dependent on past natural gas volatility and inventory.

In conclusion, the study bridges the methodological divide between traditional econometric models and contemporary machine learning techniques, offering a novel perspective on energy market volatility forecasting. The findings support a balanced approach that utilizes the strengths of both methodologies, proposing a hybrid modeling framework for future forecasting in the energy sector. As the global energy market conditions are driven by technological advancements, geopolitical shifts, and environmental considerations, this research provides a baseline for future research, where a lot of other variables and more complex models can be used to further improve the predictability of energy market volatility, incorporating out-of-sample hedging and risk management aspects.  

\clearpage
\bibliographystyle{elsarticle-harv} 
\bibliography{cas-refs}

\begin{thebibliography}{43}
\expandafter\ifx\csname natexlab\endcsname\relax\def\natexlab#1{#1}\fi
\providecommand{\url}[1]{\texttt{#1}}
\providecommand{\href}[2]{#2}
\providecommand{\path}[1]{#1}
\providecommand{\DOIprefix}{doi:}
\providecommand{\ArXivprefix}{arXiv:}
\providecommand{\URLprefix}{URL: }
\providecommand{\Pubmedprefix}{pmid:}
\providecommand{\doi}[1]{\href{http://dx.doi.org/#1}{\path{#1}}}
\providecommand{\Pubmed}[1]{\href{pmid:#1}{\path{#1}}}
\providecommand{\bibinfo}[2]{#2}
\ifx\xfnm\relax \def\xfnm[#1]{\unskip,\space#1}\fi
\bibitem[{Bollerslev(1986)}]{bollerslev1986}
\bibinfo{author}{Bollerslev, T.}, \bibinfo{year}{1986}.
\newblock \bibinfo{title}{Generalized autoregressive conditional
  heteroskedasticity}.
\newblock \bibinfo{journal}{Journal of Econometrics} \bibinfo{volume}{31(3)},
  \bibinfo{pages}{307--327}.
\bibitem[{Bollerslev et~al.(1994)Bollerslev, Engle and Nelson}]{bollerslev1994}
\bibinfo{author}{Bollerslev, T.}, \bibinfo{author}{Engle, R.},
  \bibinfo{author}{Nelson, D.}, \bibinfo{year}{1994}.
\newblock \bibinfo{title}{Arch models}.
\newblock \bibinfo{journal}{Handbook of Econometrics} \bibinfo{volume}{4},
  \bibinfo{pages}{2959--3038}.
\bibitem[{Brailsford and Faff(1996)}]{brailsford1996}
\bibinfo{author}{Brailsford, T.}, \bibinfo{author}{Faff, R.},
  \bibinfo{year}{1996}.
\newblock \bibinfo{title}{An evaluation of volatility forecasting techniques}.
\newblock \bibinfo{journal}{Journal of Banking and Finance}
  \bibinfo{volume}{20}, \bibinfo{pages}{419--438}.
\bibitem[{Brown and Yücel(2002)}]{brown2002}
\bibinfo{author}{Brown, S.}, \bibinfo{author}{Yücel, M.},
  \bibinfo{year}{2002}.
\newblock \bibinfo{title}{Energy prices and aggregate economic activity: an
  interpretative survey}.
\newblock \bibinfo{journal}{The Quarterly Review of Economics and Finance}
  \bibinfo{volume}{42(2)}, \bibinfo{pages}{193--208}.
\bibitem[{De~Cian and Sue~Wing(2019)}]{decian2019}
\bibinfo{author}{De~Cian, E.}, \bibinfo{author}{Sue~Wing, I.},
  \bibinfo{year}{2019}.
\newblock \bibinfo{title}{Global energy consumption in a warming climate}.
\newblock \bibinfo{journal}{Environmental and Resource Economics}
  \bibinfo{volume}{72}, \bibinfo{pages}{365--410}.
\bibitem[{Diebold and Lopez(1996)}]{diebold1996}
\bibinfo{author}{Diebold, F.}, \bibinfo{author}{Lopez, J.},
  \bibinfo{year}{1996}.
\newblock \bibinfo{title}{8 forecast evaluation and combination}.
\newblock \bibinfo{journal}{Handbook of Statistics} \bibinfo{volume}{14},
  \bibinfo{pages}{241--268}.
\bibitem[{Dritsaki(2017)}]{dritsaki2017}
\bibinfo{author}{Dritsaki, C.}, \bibinfo{year}{2017}.
\newblock \bibinfo{title}{An empirical evaluation in garch volatility modeling:
  Evidence from the stockholm stock exchange}.
\newblock \bibinfo{journal}{Journal of Mathematical Finance}
  \bibinfo{volume}{7(2)}, \bibinfo{pages}{366--390}.
\bibitem[{Efimova and Serletis(2014)}]{efimova2014}
\bibinfo{author}{Efimova, O.}, \bibinfo{author}{Serletis, A.},
  \bibinfo{year}{2014}.
\newblock \bibinfo{title}{Energy markets volatility modelling using garch}.
\newblock \bibinfo{journal}{Energy Economics} \bibinfo{volume}{43},
  \bibinfo{pages}{264--273}.
\bibitem[{Engle(1982)}]{engle1982}
\bibinfo{author}{Engle, R.}, \bibinfo{year}{1982}.
\newblock \bibinfo{title}{Autoregressive conditional heteroscedasticity with
  estimates of the variance of united kingdom inflation}.
\newblock \bibinfo{journal}{Econometrica: Journal of the econometric society} ,
  \bibinfo{pages}{987--1007}.
\bibitem[{Engle and Kroner(1995)}]{engle1995}
\bibinfo{author}{Engle, R.}, \bibinfo{author}{Kroner, K.},
  \bibinfo{year}{1995}.
\newblock \bibinfo{title}{Multivariate simultaneous generalized arch}.
\newblock \bibinfo{journal}{Econometric theory} \bibinfo{volume}{11(1)},
  \bibinfo{pages}{122--150}.
\bibitem[{Ghoddusi et~al.(2019)Ghoddusi, Creamer and Rafizadeh}]{ghoddusi2019}
\bibinfo{author}{Ghoddusi, H.}, \bibinfo{author}{Creamer, G.},
  \bibinfo{author}{Rafizadeh, N.}, \bibinfo{year}{2019}.
\newblock \bibinfo{title}{Machine learning in energy economics and finance: A
  review}.
\newblock \bibinfo{journal}{Energy Economics} \bibinfo{volume}{81},
  \bibinfo{pages}{709--727}.
\bibitem[{Glosten et~al.(1993)Glosten, Jagannathan and Runkle}]{glosten1993}
\bibinfo{author}{Glosten, L.}, \bibinfo{author}{Jagannathan, R.},
  \bibinfo{author}{Runkle, D.}, \bibinfo{year}{1993}.
\newblock \bibinfo{title}{On the relation between the expected value and the
  volatility of the nominal excess return on stocks}.
\newblock \bibinfo{journal}{The Journal of Finance} \bibinfo{volume}{48(5)},
  \bibinfo{pages}{1779--1801}.
\bibitem[{Haigh and Holt(2002)}]{haigh2002}
\bibinfo{author}{Haigh, M.}, \bibinfo{author}{Holt, M.}, \bibinfo{year}{2002}.
\newblock \bibinfo{title}{Crack spread hedging: accounting for time‐varying
  volatility spillovers in the energy futures markets}.
\newblock \bibinfo{journal}{Journal of Applied Econometrics}
  \bibinfo{volume}{17(3)}, \bibinfo{pages}{269--289}.
\bibitem[{Hamilton(1983)}]{hamilton1983}
\bibinfo{author}{Hamilton, J.}, \bibinfo{year}{1983}.
\newblock \bibinfo{title}{Oil and the macroeconomy since world war ii}.
\newblock \bibinfo{journal}{Journal of Political Economy}
  \bibinfo{volume}{91(2)}, \bibinfo{pages}{228--248}.
\bibitem[{Hamilton(2009)}]{hamilton2009}
\bibinfo{author}{Hamilton, J.}, \bibinfo{year}{2009}.
\newblock \bibinfo{title}{Causes and consequences of the oil shock of 2007-08}.
\newblock \bibinfo{journal}{National Bureau of Economic Research}
  \bibinfo{volume}{No. w15002}.
\bibitem[{Henriques and Sadorsky(2011)}]{henriques2011}
\bibinfo{author}{Henriques, I.}, \bibinfo{author}{Sadorsky, P.},
  \bibinfo{year}{2011}.
\newblock \bibinfo{title}{The effect of oil price volatility on strategic
  investment}.
\newblock \bibinfo{journal}{Energy Economics} \bibinfo{volume}{33(1)},
  \bibinfo{pages}{79--87}.
\bibitem[{Huang et~al.(2017)Huang, Liu, Van Der~Maaten and
  Weinberger}]{huang2017}
\bibinfo{author}{Huang, G.}, \bibinfo{author}{Liu, Z.}, \bibinfo{author}{Van
  Der~Maaten, L.}, \bibinfo{author}{Weinberger, K.}, \bibinfo{year}{2017}.
\newblock \bibinfo{title}{Densely connected convolutional networks}.
\newblock \bibinfo{journal}{Proceedings of the IEEE conference on computer
  vision and pattern recognition} , \bibinfo{pages}{4700--4708}.
\bibitem[{Karali and Ramirez(2014)}]{karali2014}
\bibinfo{author}{Karali, B.}, \bibinfo{author}{Ramirez, O.},
  \bibinfo{year}{2014}.
\newblock \bibinfo{title}{Macro determinants of volatility and volatility
  spillover in energy markets}.
\newblock \bibinfo{journal}{Energy Economics} \bibinfo{volume}{46},
  \bibinfo{pages}{413--421}.
\bibitem[{Kaufmann and Connelly(2020)}]{kaufmann2020}
\bibinfo{author}{Kaufmann, R.}, \bibinfo{author}{Connelly, C.},
  \bibinfo{year}{2020}.
\newblock \bibinfo{title}{Oil price regimes and their role in price diversions
  from market fundamentals}.
\newblock \bibinfo{journal}{Nature Energy} \bibinfo{volume}{5(2)},
  \bibinfo{pages}{141--149}.
\bibitem[{Kilian(2009)}]{kilian2009}
\bibinfo{author}{Kilian, L.}, \bibinfo{year}{2009}.
\newblock \bibinfo{title}{Not all oil price shocks are alike: Disentangling
  demand and supply shocks in the crude oil market}.
\newblock \bibinfo{journal}{American Economic Review} \bibinfo{volume}{99(3)},
  \bibinfo{pages}{1053--1069}.
\bibitem[{Kilian and Park(2009)}]{kilianpark2009}
\bibinfo{author}{Kilian, L.}, \bibinfo{author}{Park, C.}, \bibinfo{year}{2009}.
\newblock \bibinfo{title}{The impact of oil price shocks on the us stock
  market}.
\newblock \bibinfo{journal}{International economic review}
  \bibinfo{volume}{50(4)}, \bibinfo{pages}{1267--1287}.
\bibitem[{Li et~al.(2021)Li, Wang, Wang, Xin, He and Zhao}]{li2021}
\bibinfo{author}{Li, Y.}, \bibinfo{author}{Wang, W.}, \bibinfo{author}{Wang,
  Y.}, \bibinfo{author}{Xin, Y.}, \bibinfo{author}{He, T.},
  \bibinfo{author}{Zhao, G.}, \bibinfo{year}{2021}.
\newblock \bibinfo{title}{A review of studies involving the effects of climate
  change on the energy consumption for building heating and cooling}.
\newblock \bibinfo{journal}{International Journal of Environmental Research and
  Public Health} \bibinfo{volume}{18(1)}, \bibinfo{pages}{40}.
\bibitem[{Lin and Tamvakis(2001)}]{lin2001}
\bibinfo{author}{Lin, S.}, \bibinfo{author}{Tamvakis, M.},
  \bibinfo{year}{2001}.
\newblock \bibinfo{title}{Spillover effects in energy futures markets}.
\newblock \bibinfo{journal}{Energy Economics} \bibinfo{volume}{23(1)},
  \bibinfo{pages}{43--56}.
\bibitem[{Ling and McAleer(2003)}]{ling2003}
\bibinfo{author}{Ling, S.}, \bibinfo{author}{McAleer, M.},
  \bibinfo{year}{2003}.
\newblock \bibinfo{title}{Asymptotic theory for a vector arma-garch model}.
\newblock \bibinfo{journal}{Econometric Theory} \bibinfo{volume}{19},
  \bibinfo{pages}{278--308}.
\bibitem[{Ljung and Box(1978)}]{ljung1978}
\bibinfo{author}{Ljung, G.}, \bibinfo{author}{Box, G.}, \bibinfo{year}{1978}.
\newblock \bibinfo{title}{On a measure of lack of fit in time series models}.
\newblock \bibinfo{journal}{Biometrika} \bibinfo{volume}{65(2)},
  \bibinfo{pages}{297--303}.
\bibitem[{Lopez(2001)}]{lopez2001}
\bibinfo{author}{Lopez, J.}, \bibinfo{year}{2001}.
\newblock \bibinfo{title}{Evaluating the predictive accuracy of volatility
  models}.
\newblock \bibinfo{journal}{Journal of Forecasting} \bibinfo{volume}{20(2)},
  \bibinfo{pages}{87--109}.
\bibitem[{Lu et~al.(2021)Lu, Ma, Ma and Zhu}]{lu2021}
\bibinfo{author}{Lu, H.}, \bibinfo{author}{Ma, X.}, \bibinfo{author}{Ma, M.},
  \bibinfo{author}{Zhu, S.}, \bibinfo{year}{2021}.
\newblock \bibinfo{title}{Energy price prediction using data-driven models: A
  decade review}.
\newblock \bibinfo{journal}{Computer Science Review}
  \bibinfo{volume}{39(100356)}.
\bibitem[{Morana(2001)}]{morana2001}
\bibinfo{author}{Morana, C.}, \bibinfo{year}{2001}.
\newblock \bibinfo{title}{A semiparametric approach to short-term oil price
  forecasting}.
\newblock \bibinfo{journal}{Energy Economics} \bibinfo{volume}{23(3)},
  \bibinfo{pages}{325--338}.
\bibitem[{Nelson(1991)}]{nelson1991}
\bibinfo{author}{Nelson, D.}, \bibinfo{year}{1991}.
\newblock \bibinfo{title}{Conditional heteroskedasticity in asset returns: A
  new approach}.
\newblock \bibinfo{journal}{Econometrica: Journal of the econometric society} ,
  \bibinfo{pages}{347--370}.
\bibitem[{Ng and Pirrong(1994)}]{ng1994}
\bibinfo{author}{Ng, V.}, \bibinfo{author}{Pirrong, S.}, \bibinfo{year}{1994}.
\newblock \bibinfo{title}{Fundamentals and volatility: Storage, spreads, and
  the dynamics of metals prices}.
\newblock \bibinfo{journal}{Journal of Business} \bibinfo{volume}{67(2)},
  \bibinfo{pages}{203--230}.
\bibitem[{Nomikos and Pouliasis(2011)}]{nomikos2011}
\bibinfo{author}{Nomikos, N.}, \bibinfo{author}{Pouliasis, P.},
  \bibinfo{year}{2011}.
\newblock \bibinfo{title}{Forecasting petroleum futures markets volatility: the
  role of regimes and market conditions}.
\newblock \bibinfo{journal}{Energy Economics} \bibinfo{volume}{33},
  \bibinfo{pages}{321--337}.
\bibitem[{Pindyck(1999)}]{pindyck1999}
\bibinfo{author}{Pindyck, R.}, \bibinfo{year}{1999}.
\newblock \bibinfo{title}{The long-run evolution of energy prices}.
\newblock \bibinfo{journal}{The Energy Journal} \bibinfo{volume}{20(2)},
  \bibinfo{pages}{1--27}.
\bibitem[{Pindyck(2004a)}]{pindyck2004b}
\bibinfo{author}{Pindyck, R.}, \bibinfo{year}{2004}a.
\newblock \bibinfo{title}{Volatility and commodity price dynamics}.
\newblock \bibinfo{journal}{Journal of Futures Markets: Futures, Options, and
  Other Derivative Products} \bibinfo{volume}{24(11)},
  \bibinfo{pages}{1029--1047}.
\bibitem[{Pindyck(2004b)}]{pindyck2004a}
\bibinfo{author}{Pindyck, R.}, \bibinfo{year}{2004}b.
\newblock \bibinfo{title}{Volatility in natural gas and oil markets}.
\newblock \bibinfo{journal}{The Journal of Energy and Development}
  \bibinfo{volume}{30(1)}, \bibinfo{pages}{1--23}.
\bibitem[{Reboredo(2011)}]{reboredo2011}
\bibinfo{author}{Reboredo, J.}, \bibinfo{year}{2011}.
\newblock \bibinfo{title}{How do crude oil prices co-move?: A copula approach}.
\newblock \bibinfo{journal}{Energy Economics} \bibinfo{volume}{33(5)},
  \bibinfo{pages}{948--955}.
\bibitem[{Reboredo and Uddin(2016)}]{reboredo2016}
\bibinfo{author}{Reboredo, J.}, \bibinfo{author}{Uddin, G.},
  \bibinfo{year}{2016}.
\newblock \bibinfo{title}{Do financial stress and policy uncertainty have an
  impact on the energy and metals markets? a quantile regression approach}.
\newblock \bibinfo{journal}{International Review of Economics and Finance}
  \bibinfo{volume}{43}, \bibinfo{pages}{284--298}.
\bibitem[{Sadorsky(1999)}]{sadorsky1999}
\bibinfo{author}{Sadorsky, P.}, \bibinfo{year}{1999}.
\newblock \bibinfo{title}{Oil price shocks and stock market activity}.
\newblock \bibinfo{journal}{Energy Economics} \bibinfo{volume}{21(5)},
  \bibinfo{pages}{449--469}.
\bibitem[{Sadorsky(2006)}]{sadorsky2006}
\bibinfo{author}{Sadorsky, P.}, \bibinfo{year}{2006}.
\newblock \bibinfo{title}{Modeling and forecasting petroleum futures
  volatility}.
\newblock \bibinfo{journal}{Energy Economics} \bibinfo{volume}{28(4)},
  \bibinfo{pages}{467--488}.
\bibitem[{Sadorsky(2012)}]{sadorsky2012}
\bibinfo{author}{Sadorsky, P.}, \bibinfo{year}{2012}.
\newblock \bibinfo{title}{Correlations and volatility spillovers between oil
  prices and the stock prices of clean energy and technology companies}.
\newblock \bibinfo{journal}{Energy Economics} \bibinfo{volume}{34(1)},
  \bibinfo{pages}{248--255}.
\bibitem[{Shapley(1953)}]{shapley1953}
\bibinfo{author}{Shapley, L.}, \bibinfo{year}{1953}.
\newblock \bibinfo{title}{A value for n-person games}.
\newblock \bibinfo{journal}{Contributions to the Theory of Games}
  \bibinfo{volume}{II AM-28}, \bibinfo{pages}{307--317}.
\bibitem[{Slade and Thille(2006)}]{slade2006}
\bibinfo{author}{Slade, M.}, \bibinfo{author}{Thille, H.},
  \bibinfo{year}{2006}.
\newblock \bibinfo{title}{Commodity spot prices: An exploratory assessment of
  market structure and forward‐trading effects}.
\newblock \bibinfo{journal}{Economica} \bibinfo{volume}{73(290)},
  \bibinfo{pages}{229--256}.
\bibitem[{Suenaga and Smith(2011)}]{suenaga2011}
\bibinfo{author}{Suenaga, H.}, \bibinfo{author}{Smith, A.},
  \bibinfo{year}{2011}.
\newblock \bibinfo{title}{Volatility dynamics and seasonality in energy prices:
  implications for crack-spread price risk}.
\newblock \bibinfo{journal}{The Energy Journal} \bibinfo{volume}{32(3)},
  \bibinfo{pages}{27--58}.
\bibitem[{Wang and Wu(2012)}]{wang2012}
\bibinfo{author}{Wang, Y.}, \bibinfo{author}{Wu, C.}, \bibinfo{year}{2012}.
\newblock \bibinfo{title}{Forecasting energy market volatility using garch
  models: Can multivariate models beat univariate models?}
\newblock \bibinfo{journal}{Energy Economics} \bibinfo{volume}{34},
  \bibinfo{pages}{2167--2181}.

\end{thebibliography}





\end{document}